\documentclass[aps,prb,twocolumn,superscriptaddress,showpacs]{revtex4-1}
\usepackage{bm}
\usepackage{graphicx}
\usepackage{graphics}
\usepackage{amsmath}
\usepackage{amsfonts}
\usepackage{amssymb}
\usepackage{color}
\usepackage{makeidx}
\usepackage{xcolor}
\usepackage{ulem}
\usepackage{natbib}
\usepackage[unicode=true, bookmarks=false,breaklinks=false,pdfborder={0 0 1},backref=false,colorlinks=false]{hyperref}
\hypersetup{hypertex}   
\usepackage{breakurl}
\hypersetup{backref,colorlinks=true,citecolor=blue,linkcolor=blue,filecolor=blue,urlcolor=blue}

\begin{document}

\newcommand{\be}   {\begin{equation}}
\newcommand{\ee}   {\end{equation}}
\newcommand{\ba}   {\begin{eqnarray}}
\newcommand{\ea}   {\end{eqnarray}}
\newcommand{\ve}  {\varepsilon}
\newcommand{\Dis} {\mbox{\scriptsize dis}}

\newcommand{\state} {\mbox{\scriptsize state}}
\newcommand{\band} {\mbox{\scriptsize band}}

\title{Coulomb charging energy of vacancy-induced states in graphene}

\author{V.\ G.\ Miranda}
\affiliation{Instituto de F\'{\i}sica, Universidade Federal Fluminense, 24210-346 Niter\'oi, RJ, Brazil}
\author{Luis G.\ G.\ V.\ Dias da Silva}
\affiliation{Instituto de F\'{\i}sica, Universidade de S\~{a}o Paulo,
C.P.\ 66318, 05315--970 S\~{a}o Paulo, SP, Brazil}
\author{C.\ H.\ Lewenkopf}
\affiliation{Instituto de F\'{\i}sica, Universidade Federal Fluminense, 24210-346 Niter\'oi, RJ, Brazil}

\date{\today}

\begin{abstract}
Vacancies in graphene have been proposed to give rise to $\pi$-like magnetism in carbon materials, a conjecture which has been supported  by recent experimental evidence. 
A key element in this ``vacancy magnetism" is the formation of magnetic moments in vacancy-induced electronic states.
In this work we compute the charging energy $U$ of a single-vacancy generated localized state for bulk graphene and graphene ribbons. We use a tight-binding  model to calculate the dependency of the charging energy $U$ on the amplitudes of the localized  wave function on the graphene lattice sites. We show that for bulk graphene $U$ scales with the system size 
$L$ as $(\ln L)^{-2}$, confirming the predictions in the literature, based on heuristic arguments. 
In contrast, we find that for realistic system sizes $U$ is of the order of eV, a value that is orders of magnitude higher than the previously reported estimates.  Finally, when edges are considered, we show that $U$ is very sensitive to the vacancy position with respect to the graphene 
flake boundaries. In the case of armchair nanoribbons, we find a strong enhancement of $U$ in certain vacancy positions as compared to the value for vacancies in bulk graphene.

\end{abstract}

%

\maketitle

\section{Introduction}
\label{sec:introduction}


Lattice defects have long been considered as an undesired presence in micro- and nanostructured devices. In many cases, they were deliberately avoided in the manufacturing processes since they modify the electronic and structural properties and are detrimental to electronic transport. More recently, this scenario has been changing as it has been shown that defects themselves can give rise to interesting physical phenomena in many condensed matter systems. 


Graphene is one of the main platforms that is contributing to the growing interest in defective nanostructures. 
In addition to the remarkable and already thoroughly explored properties of clean graphene flakes, \cite{CastroNeto09,Geim2007,roadmap} experiments in graphene with vacancies reveal a new route 
to extend the plethora of fascinating aspects of this material. \cite{Chen09,Chen2011,Nair2012,Nair2013,JustPRB2014,Brar11,Ugeda2010,Ugeda2011,EvaColapso,LopezPolinnatphys2015,Zhang2016}.

Some of the important discoveries pointed out by experiments in graphene with 
vacancies include the onset of magnetic behavior in a $p$-block system, 
\cite{Nair2012,Nair2013,JustPRB2014,Yazyev2010,Zhang2016} signatures of the Kondo effect 
in the transport properties \cite{Chen2011,Miranda14,Fritz12} and the recently reported 
atomic charge collapse.\cite{EvaColapso}

Theory predicts that a vacancy originates a midgap state pinned at $E=0$ and localized around 
the ``vacancy site". \cite{Pereira06,Pereira08,KatsnelsonPRB2011vacancies} This result is not 
exclusive to graphene and generalizes to all class of systems that can be represented by a 
bipartite-like hamiltonian with nearest neighbor interactions only 
\cite{Abrahams1994,Santhosh2011,Willians2011}. For such {systems}, a counting rule states 
that whenever an imbalance $N_{I}=|N_{A}-N_{B}|\neq 0$ between the number of 
{constituents} of the sublattices $A$ and $B$ exists, at least $N_{I}$ states pinned at zero 
energy rise. \cite{Pereira08,Abrahams1994}
\footnote{This result holds if the $A$ and $B$ sublattice constituents are of the same kind. If this is
not the case, the states are not pinned at $E=0$ \cite{Pereira08}. It worth noting that vacancy 
zero modes in graphene can also be generated even if $N_{A}=N_{B}$ as long as the system 
is composed of clusters for which $N_{A}\neq N_{B}$ inside the cluster \cite{PalaciosRossierBrey,Weik2016}.}

The picture described above relies on a simple model for the vacancy {considering} graphene $\pi$-band electrons  only. This ``$\pi$-like" magnetism in graphene with vacancies has been 
reported experimentally recently, \cite{Nair2013,Zhang2016} though its relevance for the vacancy-induced 
magnetism is a matter of a long debate in the DFT community. \cite{Palacios2012,ChiCheng2014,Nanda12,Casartelli,Francis2015,Wanderla} As we show next, 
{our results reinforce}  the arguments in favor of the relevance of $\pi$-like magnetism in graphene 
with vacancies. {Moreover, our arguments } should be also valid for $H$ adatoms, which 
{can be described } by an onsite scalar potential model. \cite{Yazyev2010,Gonzalez-Herrero2016}

A crucial aspect of the vacancy state and that is key for the understanding of the onset 
of magnetism, Kondo effect, and atomic collapse within this model is the charging energy 
of the vacancy state $U$. This parameter, also referred to as the ``Hubbard-U", encodes the local electron-electron Coulomb interaction of the vacancy state. As such, it leads to the spin splitting of the vacancy midgap state which is essential for the onset of magnetic 
behavior. \cite{Lieb89} In addition, the interplay between $U$ and the hopping between the 
midgap state into the band states sets the condition for the onset of the Kondo effect. 
\cite{Miranda14,Cazalilla12,Fritz12}
Moreover, it has been argued that the strength of $U$ controls the vacancy charge and thus the critical coupling that   
determines the appearance of the atomic collapse \cite{EvaColapso,Uchoa2016}.
 
The central focus of this work is the determination of the charging energy $U$ of vacancies in graphene.  We address this issue for bulk graphene as well as for graphene with edges. 
Previous analytical results {on the} wave function of a vacancy in bulk graphene 
\cite{Pereira06,Pereira08,Nanda12} and graphene with edges \cite{Deng14,Deng15} 
are extremely relevant for the results we present.

We recall that, for a single vacancy in bulk graphene, the vacancy wave function decays 
as $\sim 1/R$ with $R$ being the distance from the vacancy site. \cite{Pereira06,Pereira08,Nanda12} 
This unusual behavior produces a non-normalizable wave function and hence the state is 
said to be quasi-localized with the degree of localization of the system decaying as the 
system enlarges. {This feature has led} some authors to claim this would make $U$ negligible 
for real samples. \cite{PalaciosRossierBrey,Palacios2012}

It is well known that the presence of edges strongly modifies the electronic properties of graphene. 
\cite{Brey_Fertig,ZhengPRB2007,CaioEduAntonio,Ricardinho,Tao2011,MagdaNat2014,Fasel2015} 
When edges are introduced, important changes occur also on the vacancy state. \cite{Deng14,Deng15,PalaciosRossierBrey,Bahamonprb2010,Orlofprb2013,Kawazoe,Midtvedt2015}
{Analytical accounts} on the influence of edges in the vacancy-generated state in graphene 
{have been recently put forward in Refs. \onlinecite{Deng14,Deng15}.} In {these} 
works, Deng and Wakabayashi studied the influence of edges on the vacancy localized 
state of a semi-infinite sheet\cite{Deng14} and graphene nanoribbons. \cite{Deng15} 
{Interestingly,} for the semi-infinite systems in the presence of an armchair edge,
 the vacancy state {decays} as $1/R^2$ and is thus normalizable. \cite{Deng14} 
 When a zigzag edge is considered, the influence on the vacancy state depends on which 
 sublattice the vacancy is created. {Deng and Wakabayashi} \cite{Deng14} find that if the 
 vacancy and the edge sites belong to different sublattices, the vacancy causes no effect on the 
 system zero energy states. In contrast, when both, edge and vacancy belong to the same 
 sublattice, the vacancy {yields a strongly distorted, non-normalizable} zero-mode wave 
 function around the vacancy {site}. \cite{Deng14}  
{The case of vacancies in graphene nanoribbons was studied by the same authors in 
Ref. \onlinecite{Deng15}. They } claim that a single-vacancy has no effect on the 
{zero-energy} states of zigzag-terminated edge ribbons, quantum dots, and armchair-terminated 
metallic ribbons. However, for armchair-terminated {semiconductor} ribbons, the vacancy 
generates a square-normalizable wave function pinned at $E=0$. 

Although the main focus of the present paper is the analysis of the charging energy of the 
vacancy-generated state in bulk graphene and graphene with edges, we also perform a 
systematic study of the vacancy-generated wave functions for graphene with edges. 
We combine analytical and numerical techniques to study how the properties of the 
vacancy state and its respective $U$ is altered for different system sizes, edges and 
vacancy-edge-sites distance. 

One of the most important results of this work is the derivation of an analytical expression 
for the computation of $U$. This result reveals that {in addition to the \textit{inter-site} 
contribution to the Coulomb charging at the vacancy, there is also a sizable \textit{intra-site} 
contribution which has been} so far overlooked. We note that, 
for bulk graphene, Ref.~\onlinecite{Cazalilla12} finds that the $1/R$ decay of the 
midgap state {leads to} $U\sim \left( 2\pi \ln L\right)^{-2}$, {with} $L$ being the 
linear system size. Based on this scaling behavior, the authors estimate $U\sim 1$~meV 
for typical micron flake sizes. Here we confirm the $U$ scaling predicted in 
Ref.~\onlinecite{Cazalilla12}, however our $U$ estimates are orders of 
magnitude larger than their predictions and in line with RPA/Hubbard model 
calculations. \cite{Wehling2013,PalaciosRossierBrey} 

Very recently, scanning tunneling spectroscopy (STS) experiments\cite{Zhang2016} in graphene 
deposited on a Rh foil  report the appearance of spin-split states near vacancy sites, which is 
consistent with the presence of strong onsite interactions. The STS data show splittings of 
about 20 - 60 meV, which can, in principle, be directly compared to the effective $U$ in their 
system (which depends on the effective dielectric constant at the vacancy site).

For graphene ribbons, $U$ calculations have been performed for the case of armchair ribbons in 
Ref.~\onlinecite{PalaciosRossierBrey}. The authors find out that $U$ is related to the inverse 
participation ratio (IPR) of the vacancy state, in agreement with the results we present below. 
However, {their} prediction that $U$ vanishes for increasing ribbon widths \cite{PalaciosRossierBrey} 
is not supported by our study. 
In this paper, we also perform a systematic study of the of the IPR and $U$ of the vacancy states 
for different ribbon widths and edge-vacancy-sites distances, a study lacking in the literature 
{to the best of our knowledge.}
Our results point that for some vacancy-edge-sites distances, $U$ decreases with increasing 
ribbon widths and approaches the bulk estimates, however remaining close to $\sim 1$eV even 
for real sample sizes, contrary to the vanishing $U$ predicted elsewhere. \cite{Palacios2012,Cazalilla12} 
We also find that there are some vacancy-edge-sites distance configurations for which a 
directionality effect of the vacancy wave function makes the IPR and $U$ system-size independent, 
being a truly localized state. This is the second main contribution of our work.

The paper is structured as follows. In Sec.~\ref{sec:U}, we present the tight-binding formalism 
used to obtain the vacancy wave functions and to derive the analytical expression of $U$. 
In Sec.~\ref{sec:Ubulk}, we use the {results of} the previous section to obtain $U$ estimates 
in bulk graphene for varying system sizes. In Sec.~\ref{sec:edges}, we make a thorough study 
of the vacancy state in graphene armchair ribbons for different configurations. We calculate 
the IPR of the midgap states to quantitatively evaluate the degree of localization of such states. 
We finish the section addressing the issue of vacancies in the presence of zigzag and quantum 
dots with both zigzag and armchair edges and show that our main findings for the armchair edges 
seems to remain robust to the additional presence of zigzag edges. In Sec. \ref{sec:Uac} we 
evaluate $U$ for armchair ribbons and show that $U$ mimics the IPR behavior. Also, we show 
that our results hold irrespective if the ribbons are semiconducting or metallic. Finally, 
{we present our concluding remarks } in Sec.~\ref{sec:conclusion}.

\section{Vacancy-induced midgap states: wave function and charging energy  }
\label{sec:U}

In this section, we present a single-orbital tight-binding model description of the midgap 
states due to a single vacancy in graphene monolayer systems.
We then derive analytical expressions for the Coulomb charging energy for these 
localized states. For notation compactness, we consider the  charging energy $U$ in 
vacuum. If the graphene layer is in contact with another  medium, the calculated $U$ 
should be divided by the corresponding dielectric constant $\epsilon$.

We describe single-particle spectrum and wave function using the model Hamiltonian given by 
\be
H=H_0+V,
\ee
where $V$ accounts for a single monovacancy and $H_0$ is the pristine graphene 
tight-binding Hamiltonian,\cite{CastroNeto09} namely 
\be
\label{eq:TBgraphene}
H_0=-t\sum_{\langle i,j \rangle}\left(|i \rangle \langle j|  + {\rm H.c.} \right)
\ee
where $t \approx 2.8$ eV is the nearest neighbor hopping integral, $|i\rangle$ corresponds 
to a $p_z$ orbital placed at the $i$th site of the graphene honeycomb lattice with 
interatomic separation $a = 1.41$ \AA, $\langle \cdots \rangle$ restricts the sums to 
nearest neighbors sites.

There are several equivalent ways to account for the vacancy.\cite{Deng14,Miranda14}
For analytical calculations it is convenient to model a monovacancy placed at the site 
$v$ by an on-site potential term, namely,
\be
V = V_0 |v\rangle \langle v |
\ee
and take the limit $|V_0/t|\gg 1$. Alternatively,  one can also use
\be
\label{eq:vprime}
V' = t\sum_{\langle v, j \rangle}\left(|v\rangle \langle j | + {\rm H.c.} \right),
\ee
which corresponds to turning off the hopping terms that connect the vacancy site $v$ to 
its nearest neighbors. From the numerical point of view both models, $V$ and 
$V'$, give the same results for the low-energy single-particle properties of the system. \cite{Pereira08}
(This statement is also corroborated by the good agreement between our results
and those of Ref.~\onlinecite{Deng14}.)

The single-particle electronic properties of a system with an impurity can be analytically obtained 
from a $T$-matrix analysis.\cite{Economou06} The $T$-matrix for a Hamiltonian of the  form 
$H=H_0+V$ reads $T = V (1 - V G_0)^{-1}$, where $G_0 = (E+i\eta - H_0)^{-1}$
is the Green's function of the pristine graphene system. For $|V_0/t|\gg 1$, the $T$ matrix
reduces to \cite{Economou06} 
\be
T(E) = -\frac{|v \rangle \langle v|}{ \langle v | G_0(E) | v \rangle}.
\ee
In general, the system wave functions are  obtained from the Lippmann-Schwinger equation,
$|\psi\rangle = (1 + G_0 T) |\psi^{(0)}\rangle$, namely  
\be
\psi_E(i) = \psi_E^{(0)}(i) -  \frac{ \langle i | G_0(E) | v \rangle}{ \langle v | G_0(E) | v \rangle}\,
\psi_E^{(0)}(v) 
\ee
where $\psi^{(0)}_E(i)=\langle i| \psi^{(0)}_E\rangle$ is the unperturbed wave 
function amplitude at the site $i$, solution of $H_0 |\psi_E^{(0)}\rangle= 
E |\psi_E^{(0)}\rangle$.
The knowledge of $\langle i | G_0(E) | j \rangle$ allows one to analytically calculate
$\psi_E(i)$. 

Of particular interest is the vacancy-induced midgap state $|\psi_0\rangle$
at $E=0$. However, for these states, a careful study of the $T(E \rightarrow 0)$ behaviour 
should be taken into account since, depending on the particular geometry under interest, 
$T(E \rightarrow 0)$ diverges and the strategy outlined above to obtain the wave functions 
is not always straightforward. \cite{Deng14,Deng15} In addition, it has been found that if the 
system has a finite gap, 
as in a armchair semiconducting ribbon, the vacancy-induced midgap state becomes a truly 
bound state and the wave function is no longer given by the expression above (see, for instance, 
Refs.~\onlinecite{Deng14,Deng15} for a thorough discussion of this case).

\begin{figure}[h!]
\begin{center}
\includegraphics[width=0.9\linewidth]{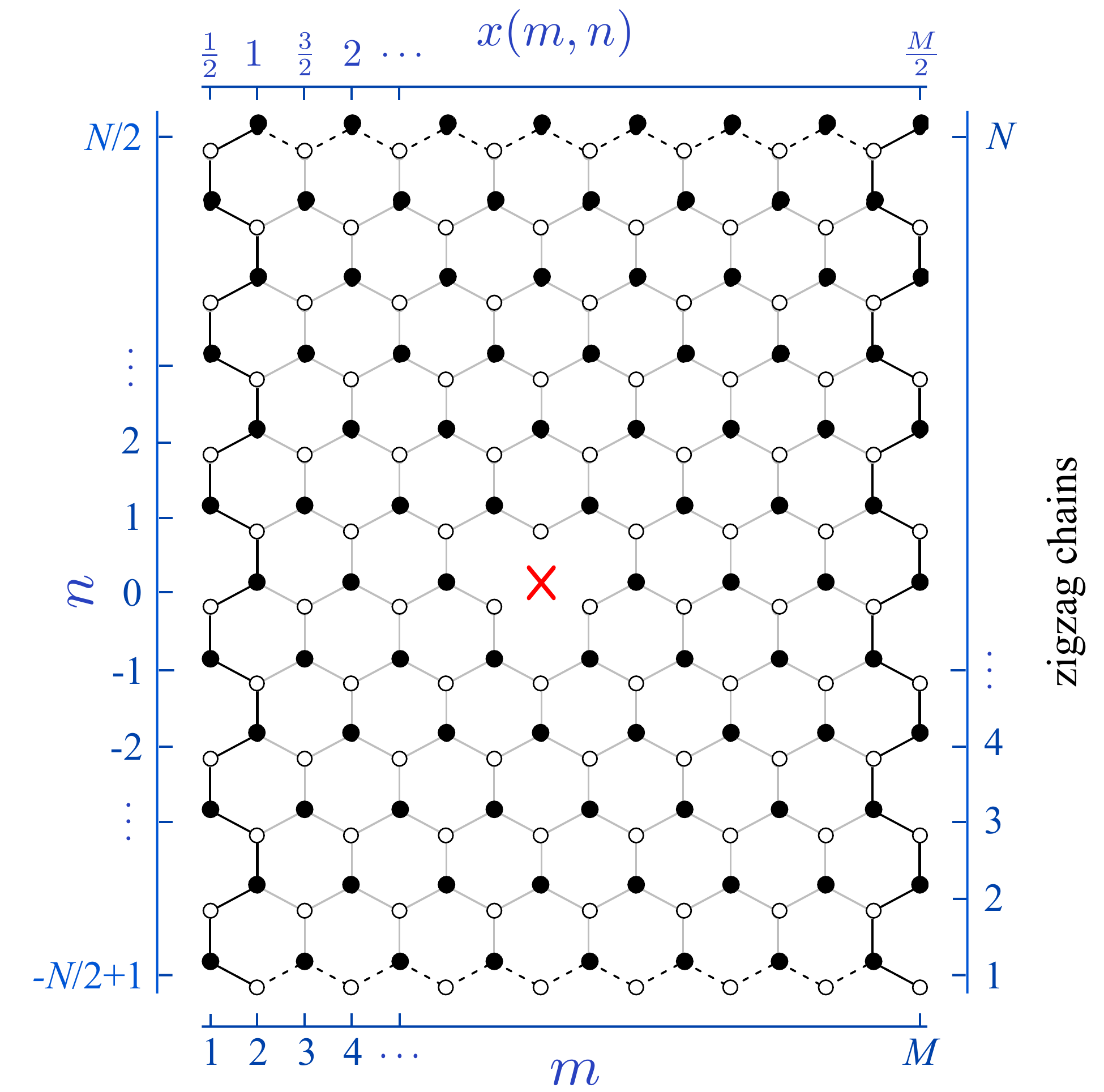}
\vskip-0.2cm
\caption{Schematics of the supercell used in this work. The supercell is defined by
$M/2$ ``vertical" armchair chains and $N$ ``horizontal" zigzag ones. $m$ is the discrete 
horizontal position with respect to the edge and $n$ labels the horizontal zigzag chains
starting from the one containing the vacancy, indicated by the cross.  }
\label{fig:lattice}
\end{center}
\end{figure}

We also analyze the midgap state problem numerically. The vacancy breaks down the lattice 
translational invariance. For bulk systems, we find it convenient to introduce 
a supercell of dimension $N_{\rm tot} = N \times M$ sites with periodic boundary conditions 
at its edges (the description of the unit cell for graphene nanoribbons is presented in 
Sec.~\ref{sec:edges}).
Figure \ref{fig:lattice} shows the superlattice geometry: the atomic lattice sites $i$ are 
denoted  by the labels $(m,n,S)$, \cite{Deng14,Deng15} where $n$ identifies the zigzag chain 
(with respect to the vacancy site) to which the site $i$ belongs and $m$ is the position of the 
$i-$th site within the chain, while $S=$ A or B is the sublattice index.   

The single-orbital graphene tight-binding Bloch basis is given by \cite{Kaxiras2003}
\be
\label{blochbasis}
\chi_{\bm{k}j}(\bm{r})= \frac{1}{\sqrt {N_{\rm sc}}}\sum_{{\bm R}^{\prime}}e^{i \bm{k}
 \cdot \bm{R}^{\prime}}\phi(\bm{r}-\bm{t}_j-\bm{R}^{\prime}),   
\ee
where the sum runs over all $N_{\rm sc}$ supercells centered at $\bm{R}^{\prime}$, 
${\bm t}_j$ gives the position of the $j$th carbon 
atom in the supercell, and $\phi({\bm r})$ is the $p_z$ orbital atomic wave function. 

We take the limit of $N_{\rm tot} =N \times M \gg 1$. Due to band folding, the $\Gamma$ point 
gives a good representation of the first Brillouin zone of the supercell. \cite{Zunger1978,Ku2010} 
For this reason, we restrict our calculation to ${\bm k} \!=\! 0$ and drop the ${\bm k}$ label 
from now on. The crystal electronic single-particle eigenstates are the solutions of 
$H\Psi_\nu (\bm{r})\!=\! E_\nu \Psi_\nu(\bm{r})$ and read 
\be
\label{eq:Psi_3D}
\Psi_\nu(\bm{r})=\sum_{i} \psi_\nu(i)\,\chi_{i} (\bm{r})\;,
\ee
where $\psi_\nu(i)$  is the $\nu$th tight-binding wave function amplitude at the $i$th site and 
$\chi_i \equiv \chi_{{\bm k}=0,i}$.  The midgap vacancy-induced state corresponds to $\nu\!=\!0$. 
For later convenience, we introduce the envelope wave functions 
\be
\psi_\nu({\bm r}_i) \equiv \frac{1}{\sqrt{\cal A}} \psi_\nu (i),
\ee
where $\cal A$ is the supercell area. This approach allows us to calculate $\psi_0({\bm r}_i)$ by direct 
diagonalization.

Let us now analyze the charging energy $U$ corresponding to the Coulomb energy associated 
with a double occupation of the midgap state $|\psi_0\rangle$. It is rather tempting to use the 
envelope wave function $\psi_{0}(\bm{r})$ to evaluate $U$, namely \cite{Cazalilla12}
\begin{equation}
\label{eq:U}
U = e^2 \! \int \!d^2r \!\int\! d ^2r^\prime \,
\frac{|\psi_{0}(\bm{r})|^2|\psi_{0}(\bm{r}^\prime)|^2} {|\bm{r} - \bm{r}^\prime|}.
\end{equation}
As shown below, this is the main contribution to the charging energy $U$ due to the $\pi$ orbitals. 
However we find another important contribution, which has been neglected so far. 

The Coulomb energy associated to the double occupation of the vacancy-induced midgap 
state reads
\be
\label{eq:U_3D}
U = e^2 \! \int \!d^3r \!\int\! d ^3r^\prime \,
\frac{|\Psi_{0}(\bm{r})|^2|\Psi_{0}(\bm{r}^\prime)|^2} {|\bm{r} - \bm{r}^\prime|},
\ee
where $\Psi_{0}(\bm{r})$ is the three dimensional wave function of the midgap state given by 
Eq.~\eqref{eq:Psi_3D}. $U$ can then be expressed in terms of the tight-binding amplitudes as
\be
\label{eq:Usum}
U = \sum_{i,j,i',j'} \psi_0^*(i)\psi_0^{}(j)\psi_0^*(i')\psi_0^{}(j') \, W_{ij,i'j'}\;,
\ee
where the sums run over all $N_{\rm tot}$ atomic sites of the supercell and 
\begin{widetext}
\be
\label{eq:Coulomb_ijmn}
W_{ij,i'j'} = e^2\int \!d^3r \!\int\! d ^3r^\prime \,
 \phi^*({\bm r}- {\bm t}_i)\phi(\bm{r}- {\bm t}_j)
\frac{1} {|\bm{r} - \bm{r}^\prime|}
\phi^*(\bm{r}^\prime- {\bm t}_{i'})\phi(\bm{r}^\prime- {\bm t}_{j'}).
\ee
Equation \eqref{eq:Usum} does not include contributions from atomic orbitals located at 
different supercells, namely, $\bm{R} \neq \bm{R}^\prime$. Since the Coulomb integral
decreases rapidly as the orbital centers are separated, one only expects significant 
inter-supercell contributions from orbitals located at the edges of neighboring 
supercells. Those correspond roughly to a fraction $1/\sqrt {N_{\rm tot}}$ of the supercell 
sites and can be safely ignored in the limit of $N_{\rm tot} \gg 1$.
\end{widetext}

There is an extensive literature on the Coulomb integral $W_{ij,i'j'}$ in the context of 
generalizing the tight-binding ideas to obtain an atomistic total-energy method
(see, for instance, Refs.~\onlinecite{Spencer2016,Harrison1981,Goringe1997,Elstner1998} 
and references therein).
The leading matrix elements \cite{Ross1950} $W_{ij,i'j'}$ correspond to an intra-atomic (on-site) 
Coulomb repulsion matrix elements, where all orbitals belong to the same atom, and to 
inter-atomic (non-local) terms, where $i\!=\!j$ is in one atom and $i'\!=\!j'$ on another.  

Accordingly, we decompose $U$ as
\be
U = U_1 + U_2,
\ee
where $U_1$ consists of intra-atomic Coulomb repulsion terms, while $U_2$ 
contains the inter-atomic ones.

Let us first consider $U_1$, namely
\be
U_1 =
e^2\! \int\! d^3 r \! \int \! d^{3}r^{\prime}\,
\frac{|\phi(\bm{r})|^{2}|\phi(\bm{r}^{\prime})|^{2}} {|\bm{r}-\bm{r}^{\prime}|}
 \sum_{i} \left|\psi_0(i)\right|^{4} \; , 
\ee
where the Coulomb integral
\be
\label{eq:Uorbital}
U_{\rm orbital}=e^{2}\int d^3 r \!\int{d^{3}r^{\prime}\,\frac{|\phi(\bm{r})|^{2}|\phi(\bm{r}^{\prime})|^{2}}
{|\bm{r}-\bm{r}^{\prime}|}}, 
\ee
can be evaluated, for instance, by an expansion of $|\bm{r}-\bm{r}^{\prime}|^{-1}$ in 
spherical harmonics. Equation \eqref{eq:Uorbital} represents the Hartree contribution 
to $U$. In the literature $U_{\rm orbital}$ was estimated to be $\sim\! 17$ eV for free
standing graphene \cite{Ross1950,Wehling2011} and, if screening effects from electrons 
of bands other than the $\pi$ are taken into account, $U_{\rm orbital}$ reduces to 
$\sim\!8.5$ eV. \cite{Wehling2011}

Hence
\be
\label{eq:U_1}
U_1 = U_{\rm orbital} \sum_{i}\left| \psi_0(i)\right|^{4}.
\ee

Let us now address $U_2$, which is more conveniently expressed  by changing the 
integration variables as $\bm{r}\rightarrow\bm{r}-\bm{t}_i$ and 
$\bm{r}^{\prime}\rightarrow\bm{r}^{\prime}-\bm{t}_j$, namely
\be
\label{eq:Utb}
U_2 = e^{2}\sum_{i \neq j}\left| \psi_0(i) \psi_0(j) \right|^{2}
\int{d^{3}r}\!\int{d^{3}r^{\prime}\,\frac{|\phi(\bm{r})|^{2}|\phi(\bm{r}^{\prime})|^{2}}
{|\bm{r}-\bm{r}^{\prime}+\bm{t}_i-\bm{t}_j|}}.
\ee

We write
\begin{align}
\label{eq:Uoff}
U_2 = & \, e^{2}\sum_{i \neq j}\dfrac{\left| \psi_0(i) \psi_0(j)\right|^{2}}{|\bm{t}_i-\bm{t}_j|} 
\nonumber \\ 
 \times \! & \int \!d^3 r \! \int \!d^3 r^\prime \dfrac{|\phi(\bm{r})|^{2}|\phi(\bm{r}^{\prime})|^{2}}
 {\sqrt{1+\dfrac{|\delta \bm{r}|^2}{|\bm{t}_i-\bm{t}_j|^2}+
 \dfrac{2\delta \bm{r}\cdot (\bm{t}_i-\bm{t}_j)}{|\bm{t}_i-\bm{t}_j|^2}}},
\end{align}
where $\delta {\bm r} =  {\bm r} - {\bm r}'$.
The orbital wave functions amplitudes $\phi(\bm{r})$ decay quickly 
for $r/a \agt 1$ and more so the overlaps of the wave functions evaluated at distances 
$|{\bm r} - {\bm r}'|/a \agt 1$.
These observations suggest that $U_2$ can be approximated by the lowest order Taylor 
expansion in powers of $|\delta \bm{r}|/|\delta \bm{t}|$  of the square root at the r.h.s.~of 
Eq.~\eqref{eq:Uoff}. This can be checked quantitatively by comparing with the exact values 
of the integral in Eq.~\eqref{eq:Utb}. \cite{Ross1950} We find that our approximation 
overestimates $U_2$ by about $\approx 10 \%$, giving us confidence in the procedure. 
\footnote{Our approximation is also consistent with the ratios between $U_{01}$, $U_{02}$, 
and $U_{03}$ (bare) reported in Ref.~\onlinecite{Wehling2011}.} 
Hence,
\begin{align}
\label{eq:U_2}
U_2\approx & \, e^{2}\sum_{i \neq j}
\frac{ |\psi_0(i)|^2 |\psi_0(j)|^{2} }{|\bm{t}_i - \bm{t}_j|}
\int \!d^3 r \!\int\! d^3 r^{\prime}\,|\phi(\bm{r})|^{2}|\phi(\bm{r^{\prime}})|^{2} 
\nonumber \\
 \approx &\, e^{2}\sum_{i \neq j}\dfrac{ |\psi_0(i)|^2 |\psi_0(j)|^{2}}{|\bm{t}_i - \bm{t}_j|} \;,
\end{align}
since the orbitals are normalized.

In the continuum limit, the sums in Eq. \eqref{eq:U_2} can be changed 
{to} integrals and the site amplitudes can be replaced by an envelope wave 
function $\psi_{0}(\bm r)$, leading to
\be
\label{eq:U_2continuum}
U_2  \approx e^2 \! \int \!d^2r \!\int\! d ^2r^\prime \,\frac{|\psi_{0}(\bm{r})|^2|\psi_{0}(\bm{r}^\prime)|^2}
{|\bm{r} - \bm{r}^\prime|} \; .
\ee

Many-body corrections to intra and inter atomic Coulomb repulsion terms
of Eq.\ (\ref{eq:Coulomb_ijmn})
have been calculated in Ref.~\onlinecite{Wehling2011}. The latter uses 
Wannier-like orbitals  projected in the $p_z$ bands. This offers the advantage of separating the 
$p_z$ contribution to the charge screening thereby accounting for the effective partial 
two-dimensional screening expected for the electrons in graphene.

In summary, the general expression for the Coulomb repulsion term $U$ is given in terms of a 
three dimensional integral given by Eq.~\eqref{eq:U_3D}. We show how the latter can be cast into 
a two dimensional form proposed in the literature. \cite{Cazalilla12} We also find an
additional significant contribution to $U$ that is proportional to $U_{\rm orbital}$, given by
Eq.\ \eqref{eq:U_1}.

\section{$U$ of vacancy-induced localized states in bulk graphene}
\label{sec:Ubulk}

In this Section, we study the dependence of $U$ with the system size. 
In order to reach our goal, we recall that the wave function $\psi_{0}({\bm r})$ of a localized 
state due to a single-vacancy in a clean bulk graphene monolayer reads \cite{Nanda12}
\begin{align}
\label{wfcNanda}
 \psi_{0}({\bm r})=\frac{{\cal N}}{r}&\,
 \sin\!\left[(\bm{K}-\bm{K}^{\prime})\cdot \frac{\bm{r}}{2}-\theta_{\bm r}\right]
 \nonumber\\ &\,\times
 \cos\!\left[(\bm{K}+\bm{K}^{\prime})\cdot \frac{\bm{r}}{2}-\frac{\pi}{3} \right],
\end{align}
where ${\bm r}=(x,y)$ is a coordinate vector with origin at the vacancy site, 
${\cal N}$ is the normalization constant, $\bm{K}=2\pi/(3\sqrt{3}a)(-1,\sqrt{3})$ 
and $\bm{K}^{\prime}=2\pi/(3\sqrt{3}a)(1,\sqrt{3})$ denote the two inequivalent 
Dirac points in the first Brillouin zone,  and $\theta_{\bm r}=\arctan(x/y)$.
Here, we consider finite systems with $N_{\rm tot}= L \times L$ sites.
Having an analytical expression for  $\psi_{0}({\bm r})$ is key for the study of systems
with $L\agt 10^2$ (or larger than $10^4$ sites), since the computation time for exact 
diagonalization scales with $L^6$.

Pereira and collaborators \cite{Pereira08} showed that for bulk graphene 
\be
\sum_{i}\left| \psi_0(i)\right|^{4} \propto \frac{1}{(\ln L)^2},
\ee
and thus 
\be
\label{eq:Udiag}
U_1 \propto \frac{U_{\rm orbital}} {(\ln L)^2 }.
\ee

Similarly, a rough estimate of $U_2$ can be obtained \cite{Cazalilla12} by using the 
approximation $\psi_0({\bm r}) \approx {\cal N}/r$. By normalizing the envelope wave function, the
charging energy given by Eq.~\eqref{eq:U_2continuum} reads $U_2 \propto e^2/\epsilon  (\ln L)^{-2}$.

We use $\psi_{0}({\bm r})$ given by Eq.~\eqref{wfcNanda} to estimate the charging energy 
$U$ of the vacancy-induced state as a function of $L$. 
We insert the lattice wave function amplitudes $\psi_0(i) = \sqrt{{\cal A}}\, \psi_0({\bf r}_i)$ 
in Eqs.~\eqref{eq:U_1}  and \eqref{eq:U_2} to 
numerically obtain $U_1$ and $U_2$, respectively. 
To account for the effect of the substrate, we use $\epsilon =4$, {which is} consistent with the 
value measured for graphene deposited on SiO$_2$. \cite{EvaPRL2014}

\begin{figure}[h!]
\begin{center}
\includegraphics[width=0.95\columnwidth]{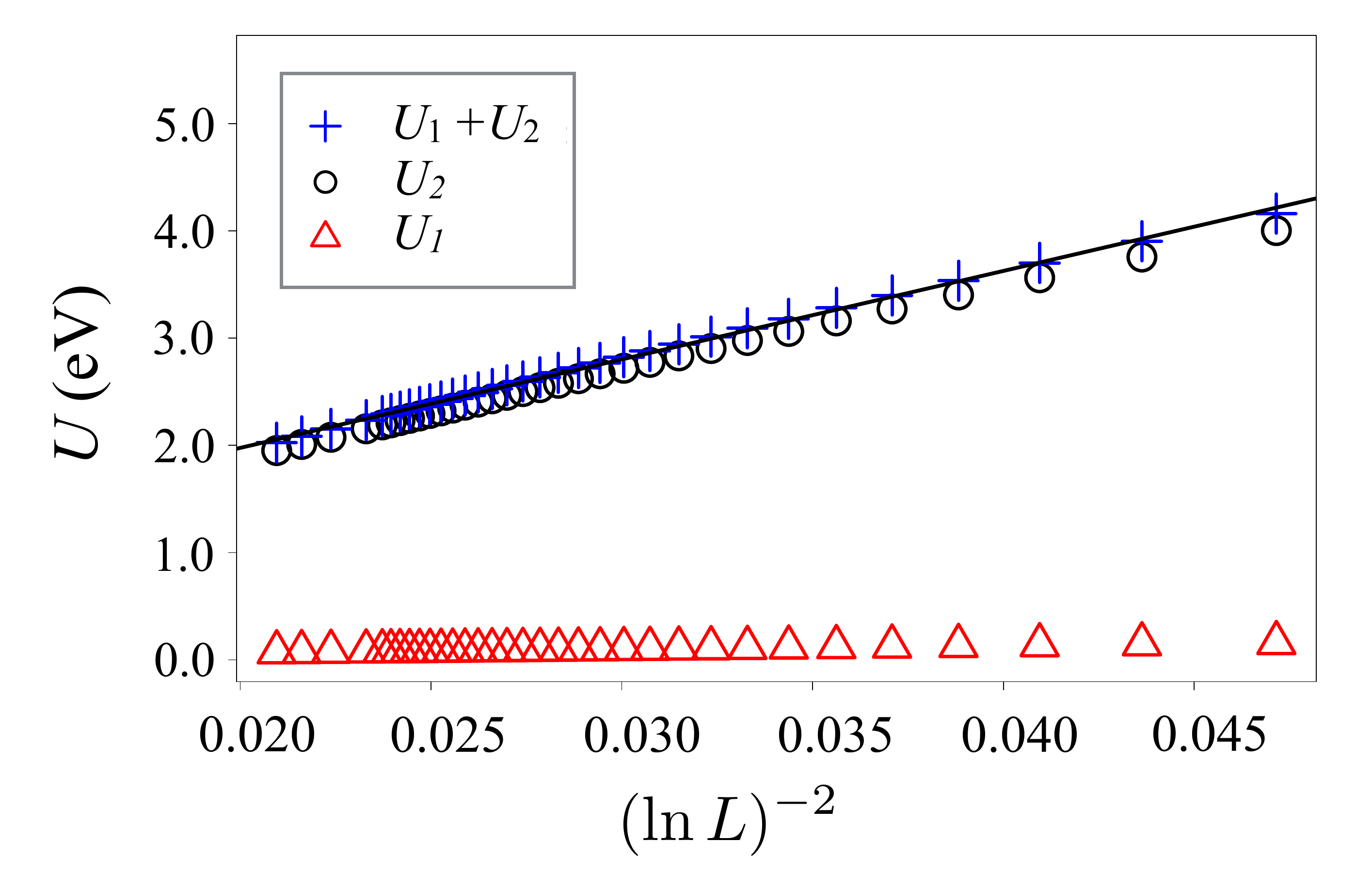}
\vskip-0.2cm
\caption{(Color online) Scaling of the diagonal and off-diagonal charging terms $U_1$ and 
$U_2$ as a function of system size. Values are in eV.
\label{fig:Uestimates}}
\end{center}
\end{figure}

Figure \ref{fig:Uestimates} gives $U_1, U_2$, and $U_{\rm approx} = U_{\rm 1}+U_{\rm 2}$
as a function of $L$.
We find that the dependence of the charging energy (in eV) with system size  is accurately 
fitted by $U_{\rm approx}=0.32+82 (\ln L)^{-2}$. 
%
%
Figure \ref{fig:Uestimates} shows that the $U_{\rm 2}$ estimates are almost an 
order of magnitude larger than those {of} $U_{\rm 1}$. 

The extrapolation of our results to graphene sheets with areas of about 1 $\mu$m$^2$ 
{(corresponding to} $L\approx 10^7$ or $10^{14}$ sites), gives $U \approx 0.64$eV. 
%
%
This {value} is two orders of magnitude larger than the value predicted in 
Ref.~\onlinecite{Cazalilla12}, but consistent with the impurity splitting predicted for an 
impurity state due to a vacancy derived from the mean field Hubbard model 
\cite{PalaciosRossierBrey} as well as from DFT calculations for small lattice sizes. \cite{Fabian2015}
{Therefore,} our results show that the $\pi-$like magnetism is not negligible in model 
systems with realistic sample sizes, in line with experimental evidences. \cite{Nair2013,Zhang2016}

\section{Graphene with edges}
\label{sec:edges}

In this section we study the influence of the edges on the vacancy-induced 
states. More specifically, we present a systematic study of the degree of localization 
and the charging energy $U$ of the vacancy-induced states in armchair nanoribbons. 
We also provide a comparative analysis with the cases of vacancies in zigzag nanoribbons 
and quantum dots. The main observation is that both localization and $U$ are strongly 
dependent on the vacancy-edge distance. 

\subsection{Midgap state wave function in the presence of armchair edges}

We follow two alternative approaches to study the characteristics of the vacancy-induced 
state: one that considers the analytical expression for the vacancy state derived in 
Ref.~\onlinecite{Deng15} and a numerical one which is based on the numerical 
diagonalization of the tight binding  Hamiltonian described in Sec. \ref{sec:U} with 
armchair boundary conditions.

Let us consider, without loss of generality, a vacancy created at sublattice $\rm{A}$. 
The wave function of $| \psi_0 \rangle$ resides solely on sublattice $\rm{B}$. In the 
notation presented in Sec.~\ref{sec:U}, the expression for $\psi_{0}(m,n,S)$ presented 
in Ref.~\onlinecite{Deng15} is written as:
\begin{widetext}
\be
\label{WFAC1}
\psi_0(m,n,S)=-\sum_{r} I_{r}(n)\left[\cos \left(  \frac{  2\pi r (  x_{B}(m,n)-x_{0}(m_{0},0)  )}{M+1} \right) -
\cos \left(  \frac{    2\pi r (  x_{B}(m,n)+x_{0}(m_{0},0)  )   }{M+1} \right)\right] \frac{\delta_{S,B}}{M+1}.
\ee

\end{widetext}
We recall that $m$ and $n$ {label} a given site $i$ and the subindex $0$ stands for the 
vacancy site (see Fig.~\ref{fig:lattice}). Here, armchair edges correspond to sites with 
$m=1,2$ and $m=M-1,M$ in Fig. \ref{fig:lattice}. Following Ref.~\onlinecite{Deng15}, 
$n=0$ stands for the zigzag row where the vacancy is positioned and the index increases 
(decreases) as one moves to rows on top (bottom), see left axis of Fig. \ref{fig:lattice}. 
Finally, $x_{B}(m,n)=m/2$ (top axis of Fig. \ref{fig:lattice}) indicates the transversal 
position on the $B$ lattice site labeled by $(m,n)$.      

The function $I_{r}(n)$ is given by \cite{Deng15}
\be
\label{IRN}
I_{r}(n)=2 (-1)^{n}
\left\{ \begin{array}{ll} 
\Theta(\frac{2\pi}{3}-k_{r})[2\cos(\frac{k_{r}}{2})]^{-(n+1)}, &   n \ge 0  \\
\Theta(k_{r}-\frac{2\pi}{3})[2\cos(\frac{k_{r}}{2})]^{(1-n)},   &   n<0  ,
\end{array} \right.
\ee
where the ``wave number" $k_r$ obeys the quantization rule dictated by 
$M$, the finite number of sites along the transversal direction 
\cite{Brey_Fertig,ZhengPRB2007,Deng15} (see {lower} panel in 
Fig.~\ref{fig:lattice}):
\be
\label{kr}
k_{r}=\frac{2\pi}{M+1}r,  \quad r=1,2,\cdots,M/2, 
\ee
 where $r$ is the band index. For the case where $\mbox{mod}(M+1,3)\neq 0$, 
 \cite{Brey_Fertig,ZhengPRB2007,Deng15} the nanoribbon is semiconducting, 
 otherwise the system is metallic. In Sec.~\ref{sec:Uac}, we use these analytical 
 expressions for the computation of the charging energy. 
 
 We also analyze the effects of vacancies in armchair graphene nanoribbons by
 numerical diagonalization of the tight binding Hamiltonian $H=H_0 + V'$, see 
 Eqs.\ (\ref{eq:TBgraphene}) and (\ref{eq:vprime}). 
As before, due to the lack of translational symmetry, we consider supercell with a 
$N_{\rm tot}=M \times N$ sites. We use periodic boundary conditions at the zigzag 
chains characterized at $n=N/2$ and $n=-N/2+1$ (see Fig.\ \ref{fig:lattice}).

\begin{figure}[h!]
\begin{center}
\includegraphics[width=0.7\columnwidth]{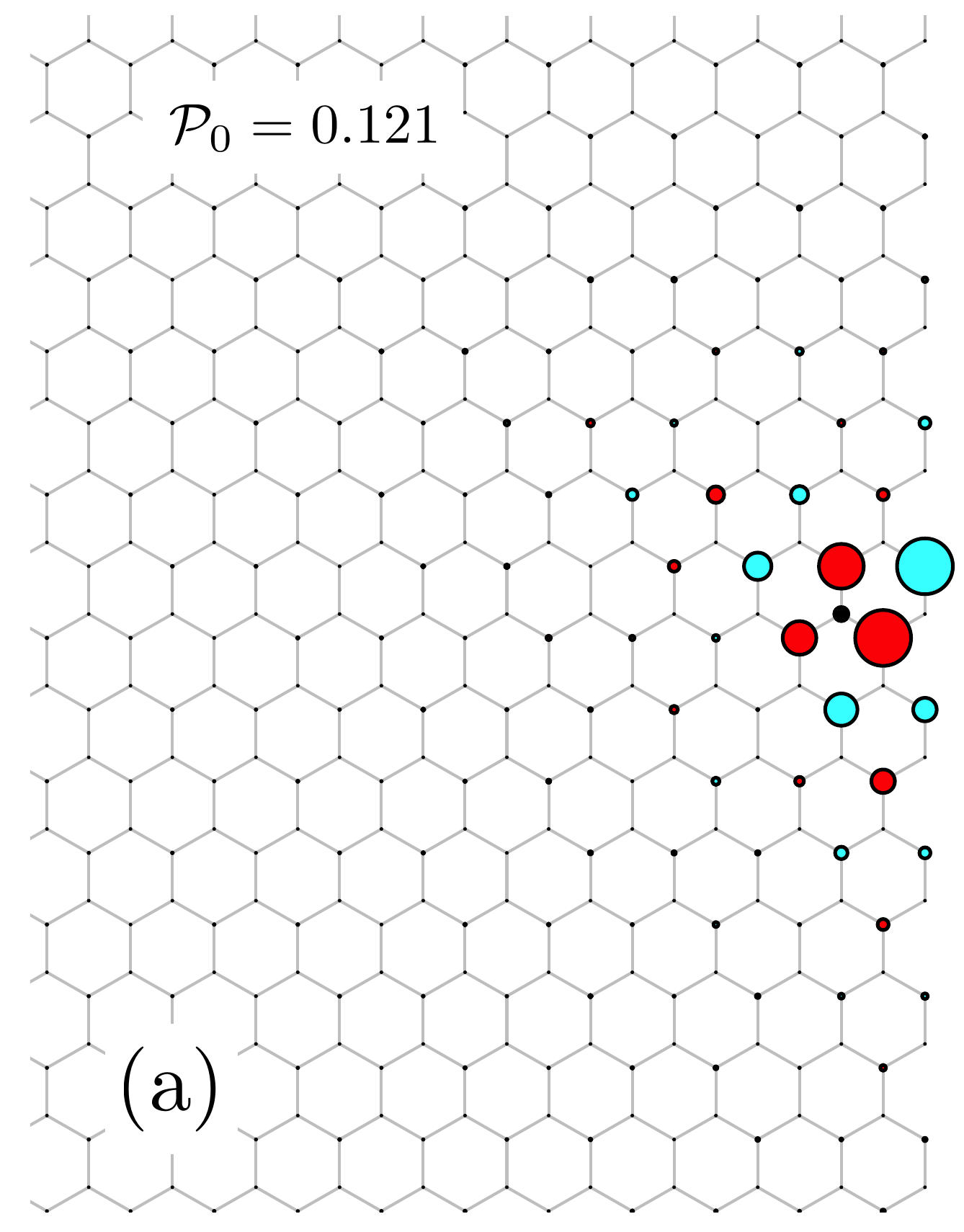}
\vskip0.3cm
\includegraphics[width=0.7\columnwidth]{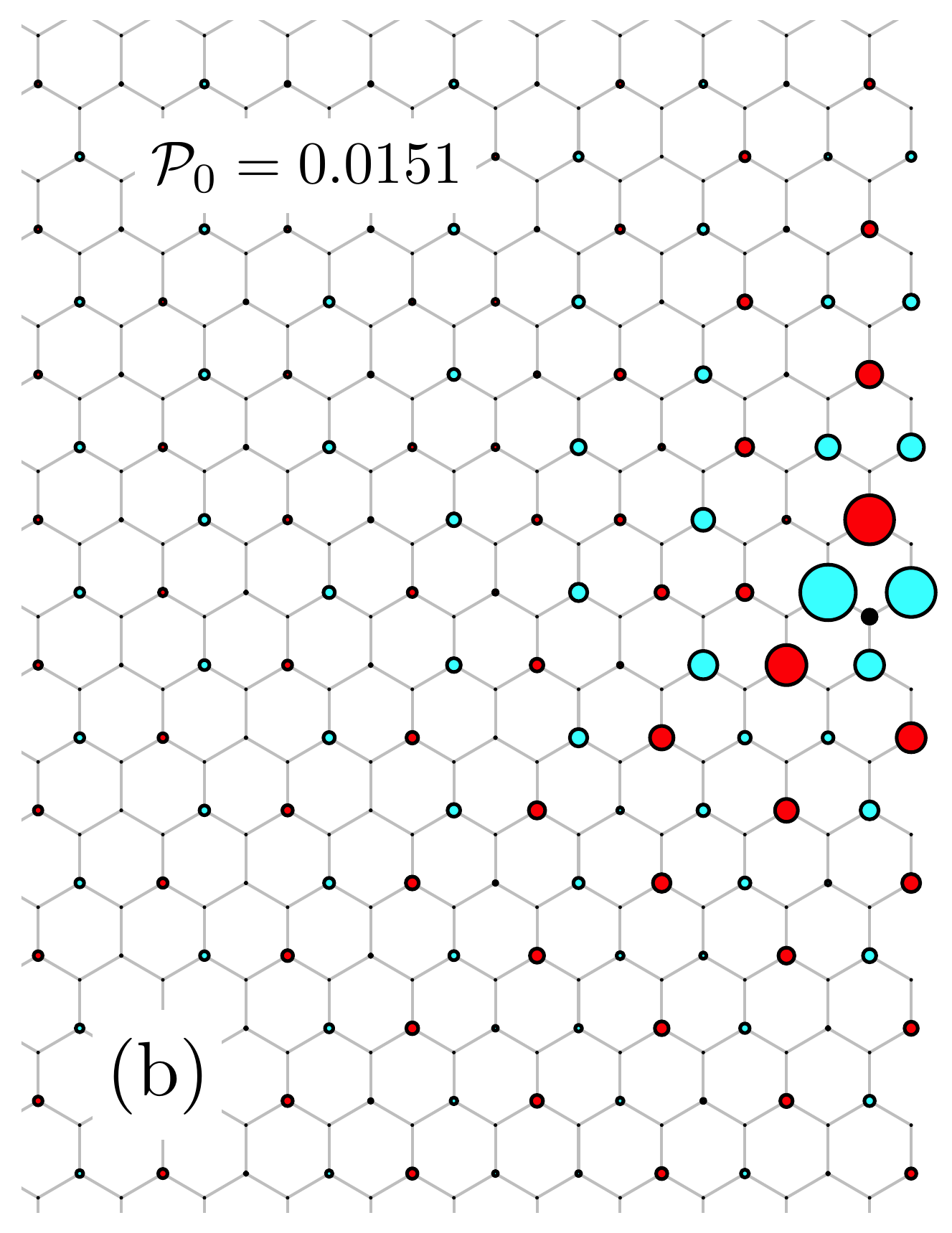}
\caption{(Color online) Wave function amplitudes for a vacancy at two different sites 
(black dots) in a armchair ribbon: (a) vacancy two sites away from the edge and 
(b) vacancy one site away from the edge. 
Note that the degree of localization drops by an order of magnitude as the vacancy 
is moved one site towards the edge. }
\label{fig:IPR}
\end{center}
\end{figure}

In Fig.~\ref{fig:IPR} we show the vacancy-generated state obtained from numerical 
diagonalization for two different vacancy-edge distances. As in the bulk, the vacancy 
state is pinned at zero energy \cite{Pereira06,Pereira08,Nanda12} and is located solely 
in the sublattice opposite to that of the vacancy. In Fig.~\ref{fig:IPR}, the black dot 
indicates the vacancy site. The radii of the bubbles are proportional to the 
amplitude of the wave functions and the blue (red) color stands for positive 
(negative) sign of the wave function. 
%
%
Note that the sites that carry the larger weights are those closer to the vacancy site. 

As expected, the vacancy originates a midgap bound state pinned at zero 
energy.\cite{Pereira06,Pereira08,Nanda12} We show in the next section that 
this numerical approach provides results which are in very good agreement 
with the analytical treatment discussed above.

%
\subsection{Midgap state degree of localization versus vacancy-edge distance.}

Although Ref. \onlinecite{PalaciosRossierBrey} presents an extensive study of the 
behavior of vacancies created in semiconducting armchair ribbons, one important aspect that was 
not properly explored so far is the dependency of the degree of localization of the vacancy state 
when the position of vacancy site is changed. Before addressing the calculation of the Coulomb 
charging energy for the vacancy-localized state, we discuss the nature of the localized wave 
functions as a function of the vacancy-edge distance $D$.

We characterize the degree of localization of the wave functions by the Inverse Participation 
Ratio (IPR) which, for a given state of energy $E_\nu$, reads:
\be
\label{eq:ipr}
\mathcal{P}_\nu=\sum_{i}|\psi_{\nu}(i)|^{4},
\ee
The IPR is contained in the interval $(0,1]$ and the closer to the upper (lower) limit, the 
higher (lower) is the degree of localization of the wave function. Note that, for the vacancy 
state, the $\mathcal{P}_{0}=U_{\rm 1}/U_{\rm orbital}$. Hence the behavior of 
$\mathcal{P}_{0}$  mimics the one followed by $U_{\rm 1}$.
 
We find that the vacancy state is extremely sensitive to the vacancy-edge distance as shown 
in Fig.~\ref{fig:IPR}. 
\footnote{On the other hand, we find that when the vacancy state is moved in the direction 
parallel the edges,the degree of localization is unchanged, which is quite reasonable since 
the system is ``infinite'' in this direction.} 
Figure \ref{fig:IPR}a shows the midgap  state for a vacancy placed two sites away from the 
edge ($D=2$, in units of $a\sqrt{3}/2$). As the vacancy is moved one position towards the 
edge ($D=1$, {as shown} in Fig.~\ref{fig:IPR}b), the vacancy state {changes abruptly} and 
the wave function extends much more than in the previous configuration. {This behavior is}
qualitatively seen {in} Fig.~\ref{fig:IPR} and quantified by the IPR. {Notice that} the IPRs of 
the two configurations differ by an order of magnitude. 

\begin{figure}[h!]
\begin{center}
\includegraphics[width=1.0\linewidth]{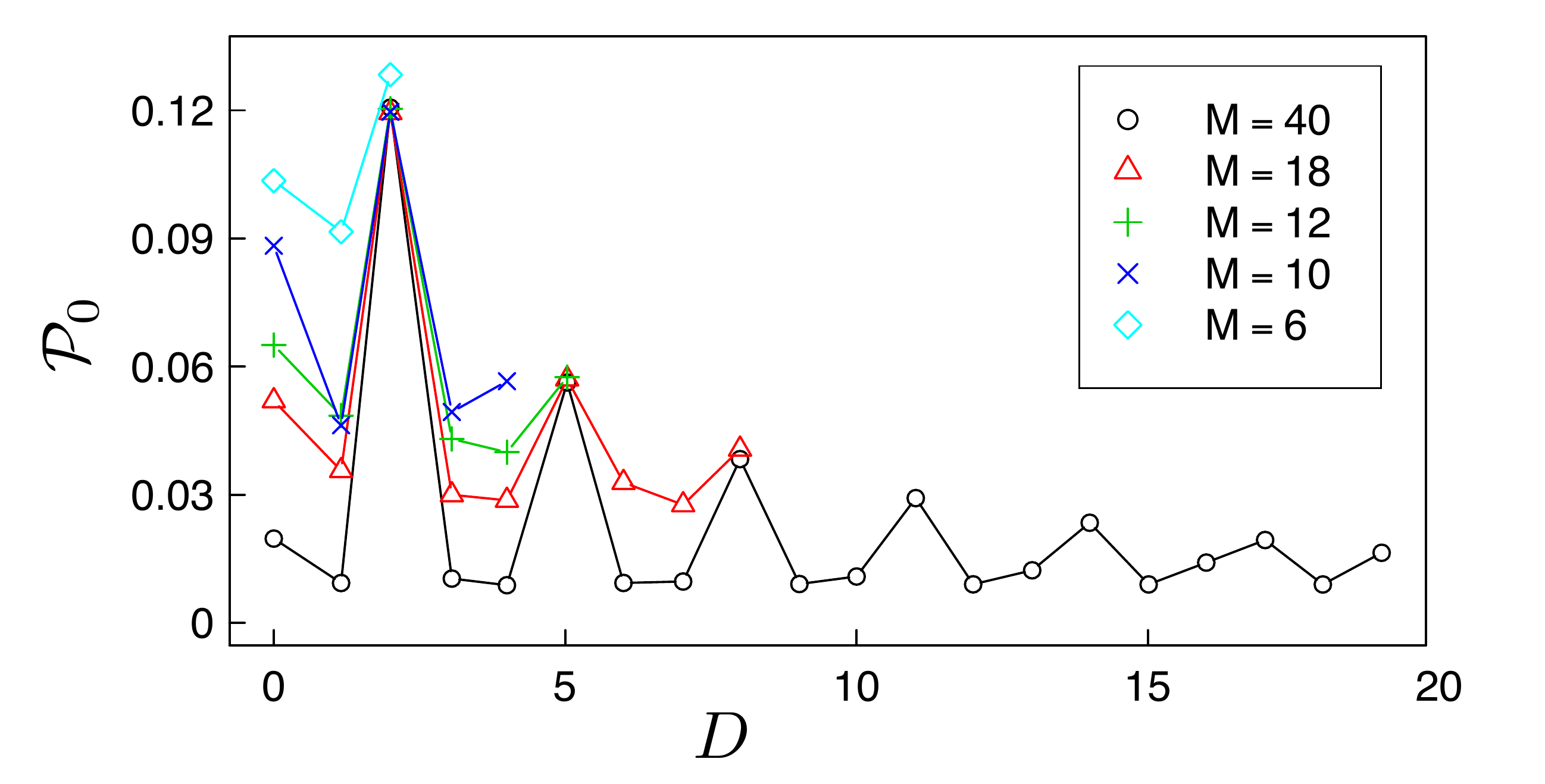}
\vskip-0.2cm
\caption{(Color online) Midgap state ${\cal P}_0$ as function as a function of $D$, the distance of the
vacancy to its closest nanoribbon edge (in units of $a\sqrt{3}/{2}$).
\label{fig:IPRSML}}
\end{center}
\end{figure}

Our numerical calculations show a non-monotonical decrease of the $\mathcal{P}_{0}$ as 
a function of the vacancy-edge distance. For very narrow ribbons, this effect is very subtle 
(see, for instance, the behavior of the IPR for the ribbon with width $M=6$ in 
Fig.~\ref{fig:IPRSML}). For small $M$, the confinement due to a finite width competes with 
the localization due to the vacancy. Figure \ref{fig:IPRSML} indicates that
the IPR does not depend on $M$ for some specific sites, while for other sites the IPR 
decreases  with increasing $M$. 
The size independent IPRs correspond to truly localized states due to the vacancy, 
while the others behave as quasi-localized states as those due to vacancies in bulk 
graphene \cite{Pereira08,Nanda12}. This interpretation plays a key role in our analysis 
and, to the best of our knowledge, {has been} unnoticed so far. 
\footnote{The latter states were probably the states observed in 
Refs.~\onlinecite{PalaciosRossierBrey,Midtvedt2015} which led them to conclude that 
going from a very narrow ribbon to bulk graphene would turn the localized state into a 
non-normalizable one with vanishing spin splitting for the midgap state. Hence, a 
vanishing magnetization.} 

The results of Fig. \ref{fig:IPR} are obtained from numerical diagonalization with $M=N=40$ 
(largest size in Fig. \ref{fig:IPRSML}), and are consistent with the analytical approach. 
Figure \ref{fig:IPRSML} shows that $\mathcal{P}_0$ becomes increasingly independent 
of system size as the ribbon width is increased. This behavior is further confirmed by the 
analysis of $U_1$ presented in Fig. \ref{fig:UvsDvarM}.

All these results are obtained for semiconducting armchair graphene nanoribbons. For metallic 
armchair ribbons we also find vacancy-induced states {which} also present the behavior observed 
in Figs.~\ref{fig:IPR} and \ref{fig:IPRSML}. This finding is at odds with the analysis of 
Ref.~\onlinecite{Deng15}. We {will} return to this discussion on Sec. \ref{sec:Uac} where we study 
the behavior of $U$ {in armchair nanoribbons.}

For semiconducting graphene nanoribbons, our findings for $\mathcal{P}_0$  can be 
qualitatively understood as follows. The combination of the bipartite nature of the honeycomb 
lattice and the presence of edges gives origin to a peculiar modulation of the degree of localization 
as a function of vacancy-edge distance. Some insight is provided from recent studies of the Kitaev 
model in the gapped phase, where it has been shown that a vacancy induces a zero mode state 
with a specific directionality such that the wave function of this state is nonzero only in a wedge emanating 
from the vacancy position and zero elsewhere. \cite{Santhosh2011,Willians2011} This directionality 
is only sublattice dependent, but independent from the site chosen within the same sublattice. 
We recall here that semiconducting armchair ribbons constitute a realization of a gapped 
honeycomb model. \cite{Brey_Fertig} Hence, we also observe a directionality pattern in the 
vacancy zero mode states in these ribbons. We note that in distinction to the Kitaev model, the systems 
we study  have edges that strongly influence the behavior of the midgap states, particularly, 
their directionality and degree of localization as a function of the vacancy site position. 

\begin{figure}[h!]
\begin{center}
\includegraphics[width=0.90\linewidth]{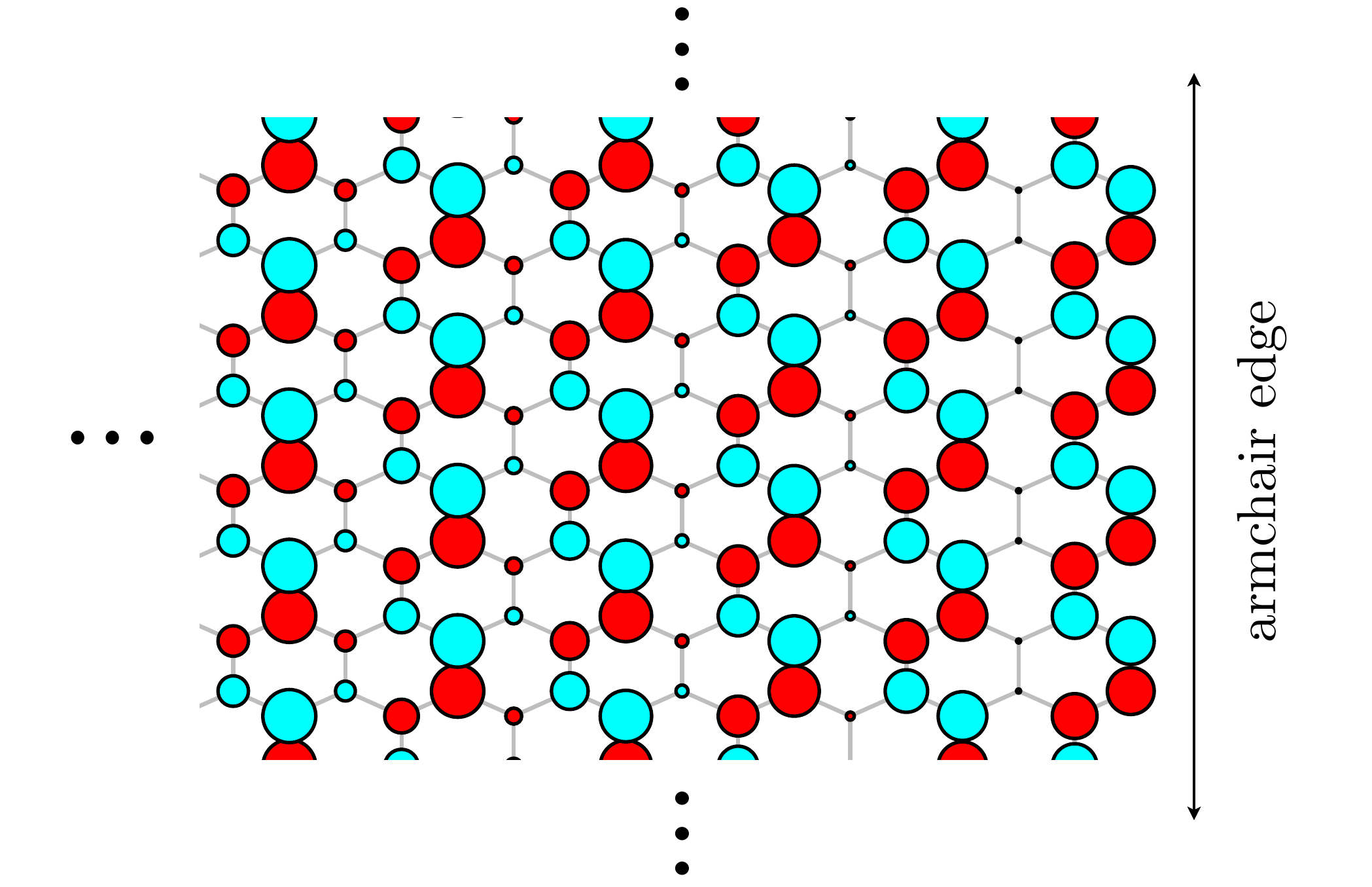}
\caption{(Color online) Site dependency of the amplitudes of the state with energy closest to zero energy 
of a pristine armchair ribbon. \cite{Brey_Fertig} The circles radii are proportional to the wave function 
amplitude, red and blue correspond the negative and positive amplitudes respectively.}
\label{fig:lowACBrey}
\end{center}
\end{figure}

Further insight is obtained by inspecting the wave function with energy closest to zero  of a clean 
armchair ribbon, illustrated in Fig.\ \ref{fig:lowACBrey}. By inspecting  the probability of finding 
the states at the sites along any of the ``horizontal'' zigzag chains starting from the armchair edge, 
see Fig.\ \ref{fig:lowACBrey}, one clearly observes a pattern of two maxima followed by a minimum. 
We note that this pattern is complementary to the one we obtain for ${\cal P}_0$ as we move the 
vacancy across a zigzag chain. Hence, if a vacancy is placed at a site where the state closest to the zero 
mode of the clean ribbon has a maximum, the vacancy state suffers a larger repulsion and has 
to extend more, the opposite occurs when the vacancy is placed at sites with vanishing weights in 
the clean system.

This view is confirmed by the study of midgap states due to hydrogen adatoms (not shown 
here) instead of vacancies. Interestingly, the manner in which the vacancy zero mode is more 
localized/extended is consequence of the directionality of the midgap state. In the sites 
where ${\cal P}_0$ {has} a maximum, the vacancy state is directed towards the closest edge 
becoming more concentrated. In the cases where a minimum in the IPR occurs, the vacancy 
state is directed to {the} farther edges 
or {to the} the ribbon's longitudinal direction and hence {becomes more extended.} 

This is an interesting property which could allow the tunability of a site-site 
entanglement \cite{Santhosh2011} of this system and also tune the transport properties 
of a specific edge due to interactions with the vacancy state. The transport along a chosen 
edge could be blocked or reduced just by selecting the vacancy/adatom position respective 
the chosen edge. As we show in Sec. \ref{sec:Uac}, all these findings have a strong impact 
on the charging energy $U$.

\subsection{Vacancies with other kinds of edges: zigzag and quantum dots}

For other kinds of edges, namely zigzag, chiral and those found in quantum dots, we also observe 
vacancy-induced states $|\psi_0 \rangle$.

\begin{figure}[h!]
\begin{center}
\includegraphics[width=0.7\linewidth]{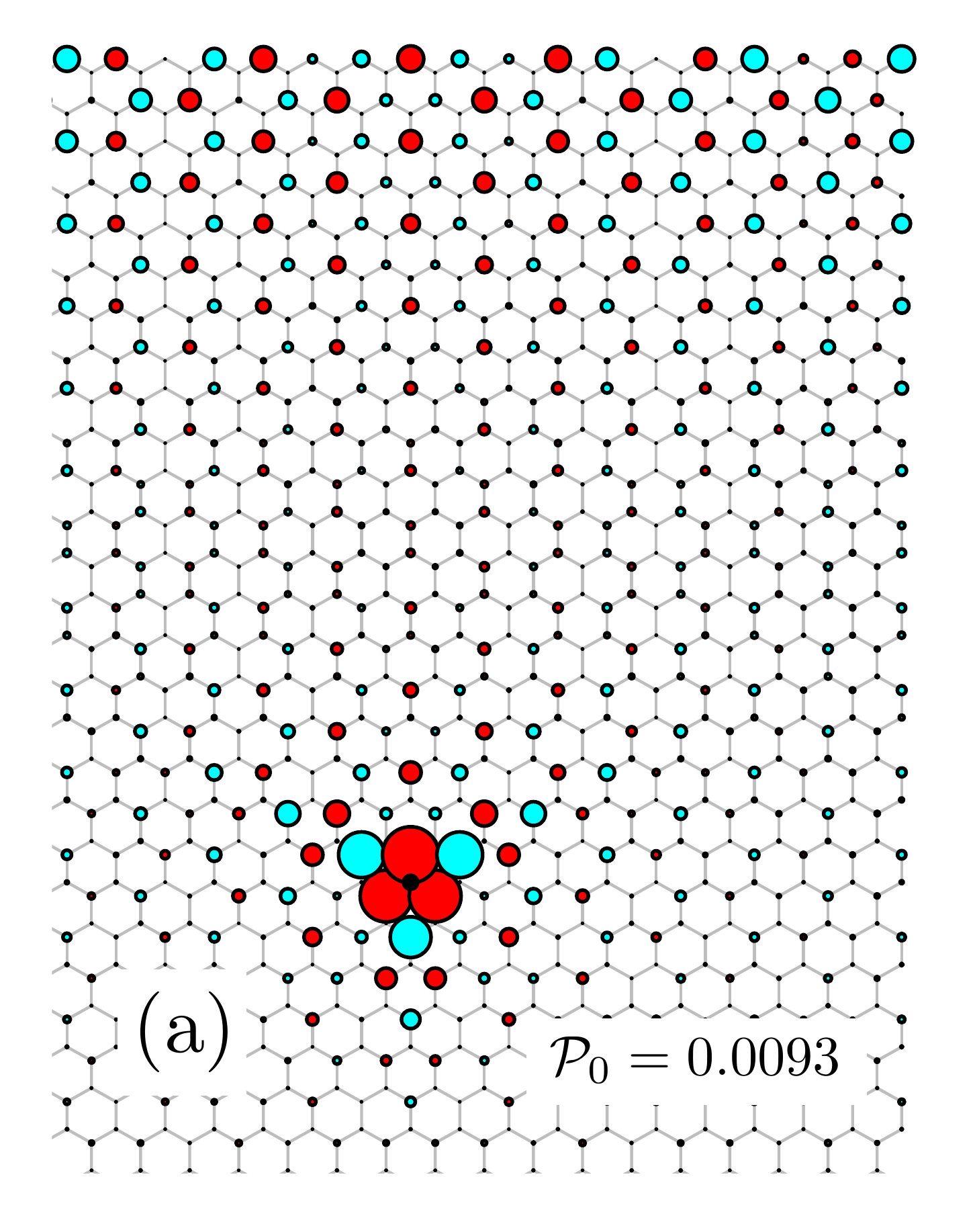}
\includegraphics[width=0.7\linewidth]{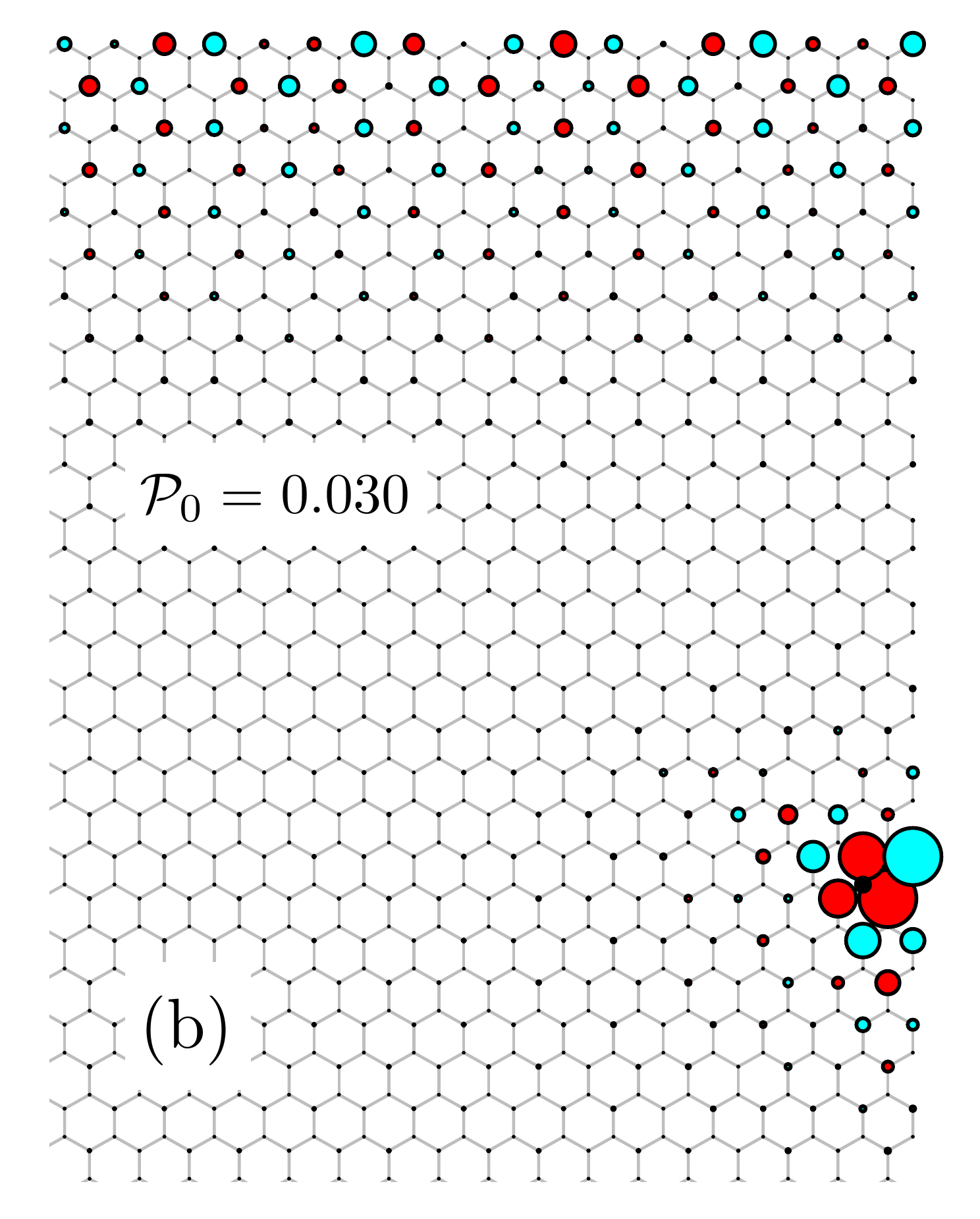}
\vskip-0.2cm
\caption{Wave function amplitudes for the vacancy state in a zigzag nanoribbon (a) 
and for a quantum dot (b). }
\label{fig:6nova}
\end{center}
\end{figure}

For zigzag ribbons,  Ref.~\onlinecite{Deng15} predicts that a vacancy does not affect 
the zero energy states of the system. Our study agrees with this result. However, we 
find that the vacancy gives rise to pairs of eletron-hole states with same energy and 
amplitude (but opposite phases), as expected from chiral symmetry preservation. 
{In addition,} $|\psi_0 \rangle$ clearly hybridizes with the edge-localized states of the edge with sites 
belonging to the opposite sublattice of the vacancy site (see Fig.\ \ref{fig:6nova}a). These observations 
are in agreement with transport simulations for this {kind} of system. \citep{Bahamonprb2010,Orlofprb2013} 

The situation becomes more complicated in the case of graphene quantum dots, whose edges are 
(in general) a combination of zigzag and armchair chains. We find that the midgap states hybridize
with the states localized at the system edges. In addition, we observe an energy shift of the vacancy 
state, {similarly to} the case of zigzag ribbons. 
Here the presence of the armchair-like edges also plays a role: We find a modulated behavior of the 
IPR of the vacancy state as the distance of the vacancy site from the armchair edge is varied. 
As in the {case of} armchair ribbons, we also observe increasing degree of localization of the 
vacancy state as the distance from the armchair edge is decreased. The IPR has its maximum 
for $D=2$ as for armchair ribbons, see Fig. \ref{fig:6nova}b.

Although we did not make an extensive numerical analysis of vacancies in zigzag ribbons or 
graphene quantum dots, the discussion above is useful to emphasize that the main results of 
our paper for the armchair ribbons should remain valid for realistic samples in which other 
kinds of edges and{/or} disorder appear. 

\section{Charging energy estimates for armchair ribbons}
\label{sec:Uac}

We focus our attention on the analysis of vacancy-induced states in armchair graphene nanoribbons 
since in this case the degree of localization of the $|\psi_0\rangle$ states is much more pronounced, 
as discussed in the previous sections. 
More specifically, we calculate the charging energy $U_1$ and $U_2$, given respectively by 
Eqs.~(\ref{eq:U_1}) and (\ref{eq:U_2}), as a function of the vacancy-edge distance $D$ and 
the nanoribbon width.

\begin{figure}[h!]
\begin{center}
\includegraphics[width=1.0\linewidth]{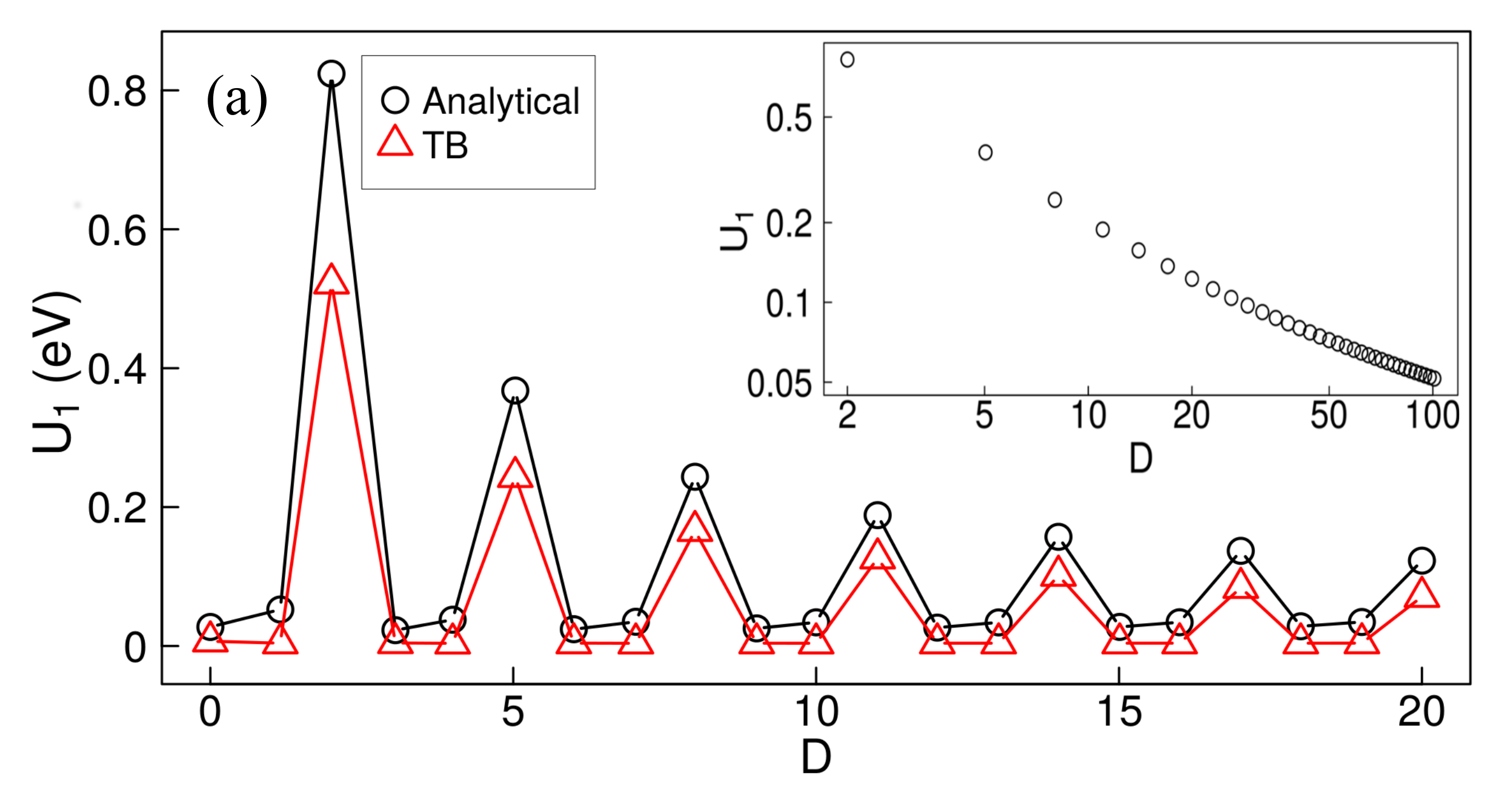}
\includegraphics[width=1.0\linewidth]{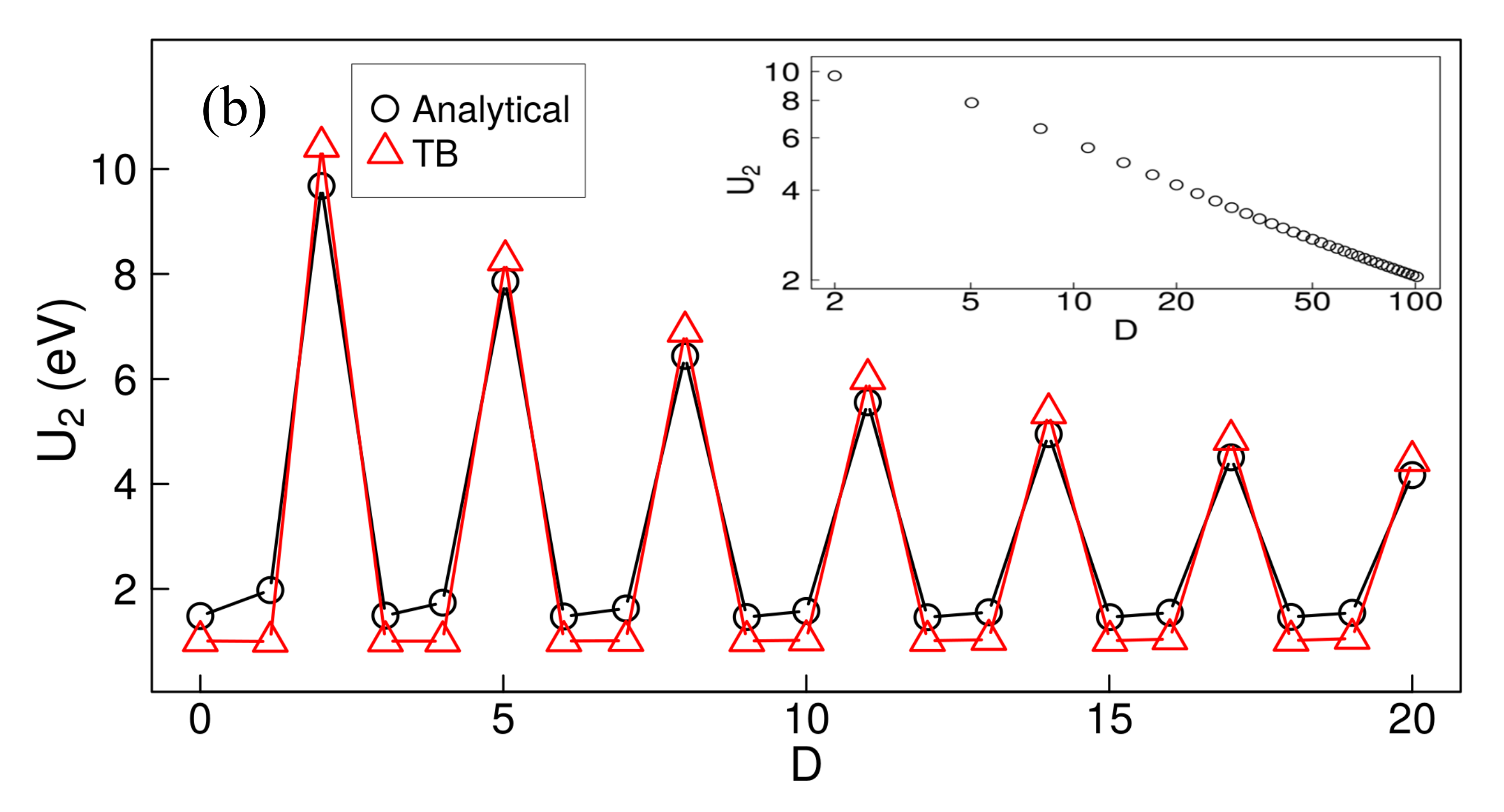}
\caption{(Color online) Charging energy $U$ as a function of the vacancy-edge distance $D$ 
for an armchair nanoribbon of width $M=220$. $U=U_1+U_2$ is calculated by using numerical 
and  analytical wave functions. Panel (a) corresponds to the intra-atomic $U_1$ contribution, 
while (b) to the inter-atomic Coulomb term $U_2$. The insets show maxima of $U_1$ and $U_2$
versus $D$ in a log-log scale.}
\label{fig:compare-Utb-Uanalytic}. 
\end{center}
\end{figure}

Both the analytical and the tight-binding approaches give the {envelope wave functions} $\psi_0(i)$
{which} allows one to numerically evaluate $U_1$ and $U_2$. We compare the results and find 
good qualitative agreement, {as shown in} 
Fig.~\ref{fig:compare-Utb-Uanalytic}. 
For the cases we considered (with $M=220$), the tight-binding values for $U_1$ are about 30\% 
smaller (for the sites with the larger $U_1$) than those obtained using the analytical wave functions. 
For $U_2$, which is one order of magnitude larger than $U_1$, both approaches agree 
within 5\%. In Appendix \ref{sec:robust}, we provide evidence that the discrepancy is not related to 
the size of the supercell. 
We speculate that the larger discrepancy in the case of $U_1$ arises from the fact that it scales 
with the numerically-obtained amplitudes as $|\psi_0(i)|^4$ (see, e.g., Eq.\ (\ref{eq:U_1})) while 
$U_2$ scales roughly as $|\psi_0(i)\psi_0(j)|^2$ with $i\neq j$ (Eq.\ (\ref{eq:U_2})).  Thus, the 
numerical values of $U_1$ tend to be more sensitive to small numerical errors in the calculation 
of the envelope wave functions $\psi_0(i)$ than those obtained for $U_2$.

The calculation of $U$ using the analytical expression for the $\psi_0(i)$ is significantly faster  and 
{demands} much less memory than the tight-binding approach, {which requires storage of the 
Hamiltonian matrix elements. This allows us to address ribbons with micron size widths. The results 
presented next are obtained using analytical wave functions.

Figure \ref{fig:UvsDvarM} shows $U_1$ and $U_2$ as a function of the vacancy-edge distance $D$. 
The results clearly indicate that $U$ is almost independent of the ribbon width $M$ for very large 
systems. Moreover, for the sites for which $U_1$ ($U_2$) has a minimum, the computed charging
energies approach the bulk values obtained in Fig. \ref{fig:Uestimates}. 
The dependence of $U_2$ {with} $D$ follows the same pattern as the IPR (for $U_1$ this is expected,
since the latter is proportional to ${\cal P}_0$), namely, an oscillatory behavior as the vacancy is moved 
away from the edge with a fast decay in the modulation (see insets of Fig.~\ref{fig:compare-Utb-Uanalytic}). 
When the vacancy is far from the ribbon edges, $U$ is suppressed with respect to its 
largest value by an order of magnitude and approaches the bulk estimate. 

\begin{figure}[h!]
\begin{center}
\includegraphics[width=1.0\linewidth]{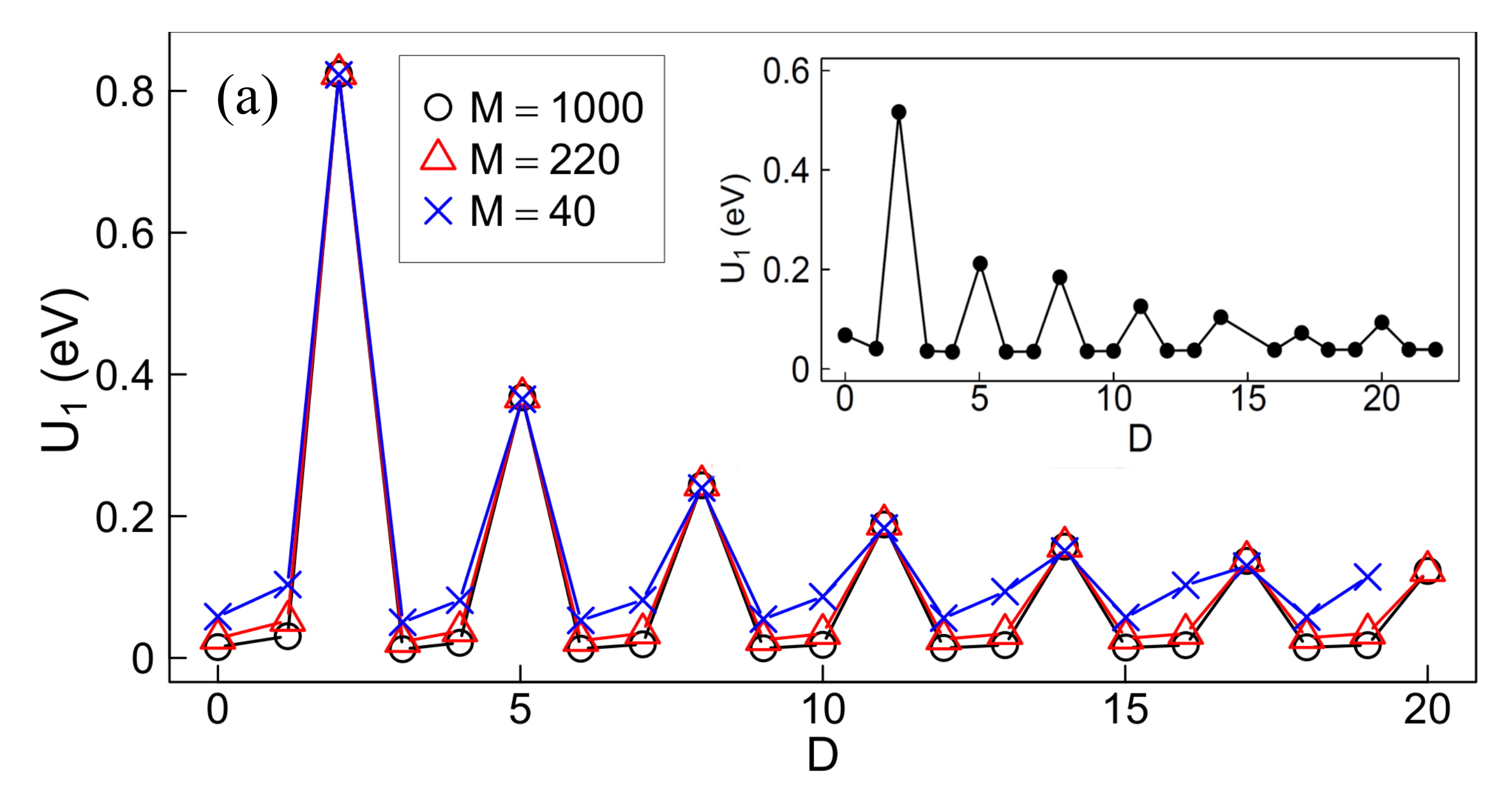}
\includegraphics[width=1.0\linewidth]{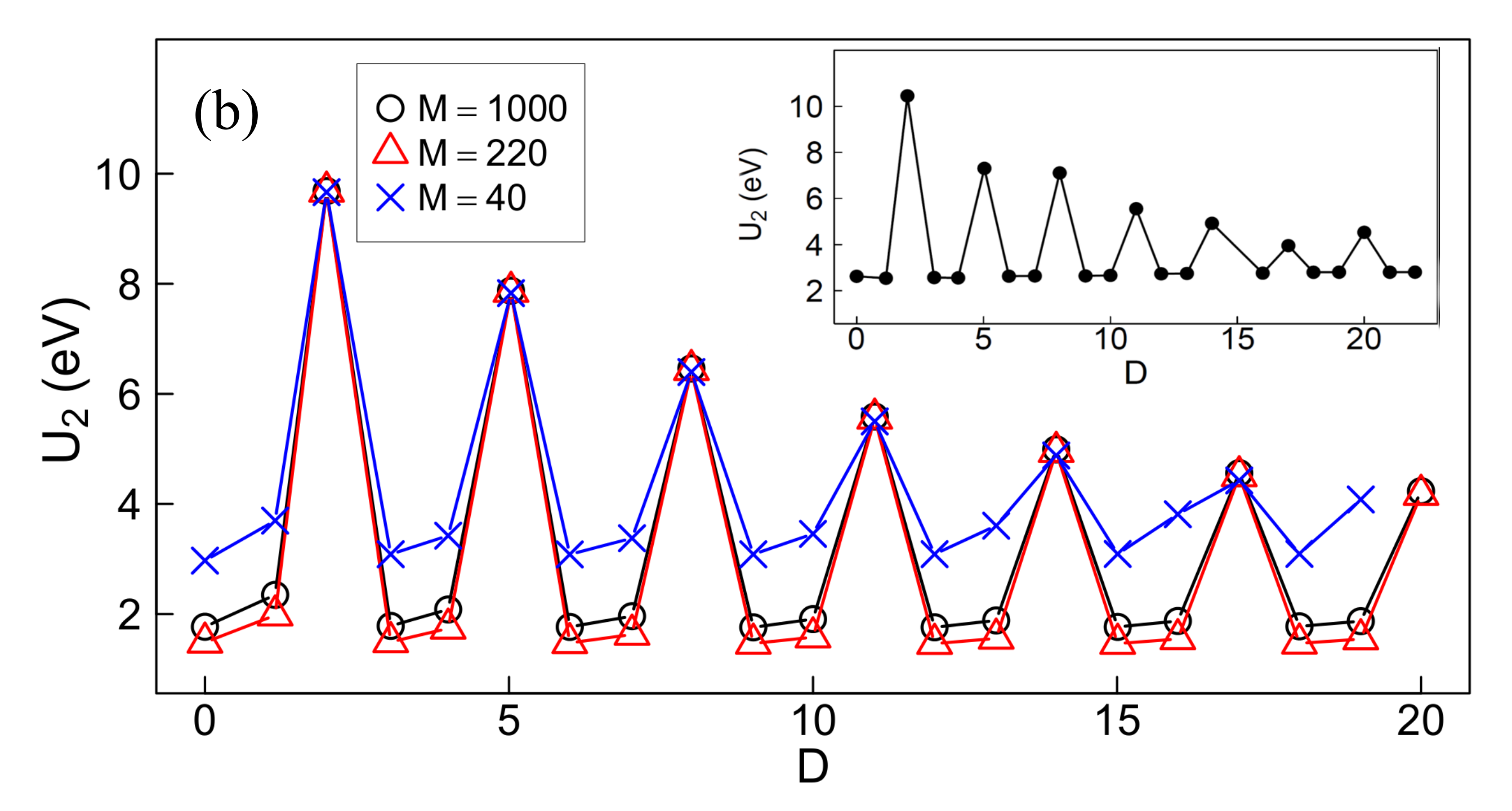}
\caption{(Color online) Charging energy $U$ as a function of the vacancy-edge distance 
$D$ for armchair nanoribbons of different widths, namely, $M=40,220,$ and 1000. The 
charging energy is split in an intra-atomic $U_1$ (a) and an inter-atomic Coulomb term 
$U_2$ (b). The insets show the tight-binding result for a metallic ribbon with $M=44$.}
\label{fig:UvsDvarM}. 
\end{center}
\end{figure}

We note that the ribbon with $M=44$ is metallic. In this case, the analytical treatment of 
Ref.~\onlinecite{Deng15} predicts that the vacancy does not produce a localized state
pinned at zero energy. Following the notation introduced in Ref.~\onlinecite {Deng15}, 
we find that, when the vacancy is placed at the so-called nodal line (sites where the 
wave function of the lowest energy state of the clean system vanishes, which correspond 
to the sites with the smallest probabilities in Fig. \ref{fig:lowACBrey}), a localized state 
pinned at zero energy {arises}, originating the maxima in the charging energy seen in the 
inset of Fig.~\ref{fig:UvsDvarM} for the width $M=44$. Interestingly, if the vacancy is
placed at a site off the nodal line, we do not observe a localized state pinned at zero 
energy, as predicted in Ref.~\onlinecite{Deng15}. 
{Instead,} we find that the vacancy originates an electron-hole pair of states close to zero 
energy with a localized character. The charging energy for these states correspond to
the ``minima" in the modulated pattern shown in the insets of Fig.~\ref{fig:UvsDvarM}. 
We note that these estimates are in excellent agreement with the behavior found for 
the semiconducting ribbons, indicating that the values for $U$ are robust irrespective to the 
metallic or semiconducting character of the ribbons. 

The appearance of localized states is metallic ribbons is important for the 
discussion of the observation of bound states immersed in the continuum (BIC) in 
graphene \cite{OrellanaEPL2010,OrellanaEPL2014} as the localized states here 
occur within a region of finite density of states.

\section{Conclusions and outlook}
\label{sec:conclusion}

In this paper we study the charging energy $U$, a key element to understand the magnetic 
properties of the system, of a localized state due to a single-vacancy in monolayer graphene bulk
and in graphene nanoribbons. 

We find that $U$ can be expressed in terms of two main contributions, $U_1$ and $U_2$, 
corresponding respectively to intrasite and intersite electron-electron interactions. We show that 
$U_2$ can be identified with the effective low energy expression for the Coulomb energy 
associated with two-dimensional electronic wave functions. 
Although $U_2$ is the dominating term, there are several scenarios where the $U_1$ contribution 
to the charging energy can become important.
For instance, a recent study \cite{Uchoa2016} suggests the use of impurities to design a lattice 
structure that can give rise to a graphene-based spin-liquid system.
There it is argued that the onset of the spin-liquid regime depends on the ratio between the charging 
energy of the impurity bound state and the hopping between these impurity states, \cite{Uchoa2016} 
that makes the precise assessment of $U$ very critical to infer the system behavior. 

Our systematic study of the charging energy confirms the heuristic prediction that $U$ scales 
with the sample length $L$ as $(\ln L)^{-2}$. Our estimates for $U$ support the picture of $\pi$ 
magnetism in realistic sample sizes,  contrary to previous results. \cite{PalaciosRossierBrey,Palacios2012}

Edges change significantly this simple bulk scaling. In this paper, we perform a systematic investigation of the midgap states 
in armchair ribbons. We establish the dependence of $U$ and the IPR with the vacancy-edge distance and 
discuss how this is related to the directionality of the wave functions. Some midgap states are truly localized 
(unnoticed so far) with IPR and $U$ an order of magnitude larger than the bulk value. Similar behavior is 
also found for metallic armchair ribbons, which could be a manifestation of a BIC. This particular case is at 
odds with the wave function analysis presented in Ref.~\onlinecite{Deng15}. For the remaining cases 
the overall agreement is very good. Moreover, we show that other kinds of edges also affect the 
vacancy-induced state. In particular, for zig-zag edges and quantum dots, the vacancy-induced states 
have energies shifted from the Dirac point due to a strong hybridization with edge-induced localized 
states. This observation can be an indication that our results for armchair edges can be robust in 
samples with edges other than zigzag and armchair. 

Using the tight-binding model, we have checked that an hydrogen adatom gives raise to localized states 
that share some of the features of those caused by a vacancy. Their IPR, for instance, have the same 
oscillatory behavior we find for the armchair ribbons with vacancies. This opens the possibility of having 
a controlled way to verify our results by the precise manipulation of H adatoms using a STM tip and to observe 
the generated states pattern by a STS measurement near the adatom, as recently shown in 
Ref.\ \onlinecite{Gonzalez-Herrero2016}. The results we present can also be checked experimentally 
in other platforms since they are valid for other systems with bipartite lattices. Artificial graphene could 
also be used to verify experimentally our findings.

We stress that a correct assessment of the substrate dielectric constant $\epsilon$ is key 
for a quantitative comparison of our $U$ estimates with the experimental values. We recall 
that the estimates we present here were obtained considering $\epsilon=4$ which is 
characteristic of graphene on top of SiO$_2$. \cite{EvaPRL2014} Depending on the 
dielectric environment to which graphene is exposed $\epsilon$ can vary by two orders 
of magnitude. \cite{Uchoareview} Hence, the dielectric media to which graphene is submitted 
in the recents experiments of  Refs.~\onlinecite{Gonzalez-Herrero2016,Zhang2016}, which 
use graphene on top of SiC and Rh substrates, respectively, can have an important 
influence in the discrepancies between our estimates and the observed $U$. A systematic 
experimental study of the role of the substrate and proper characterization of the dielectric 
constant that should be used in the $U$ estimates is necessary to clarify this issue.

Our $U$ estimates can be directly comparable to that of Ref.~\onlinecite{Cazalilla12}, which also 
estimates $U$ for graphene on SiO$_2$ substrates. We recall that our results are $2 - 3$ orders of magnitude 
larger than the values predicted in that paper. We note that our estimates are based on a rigorous derivation 
of an expression for $U$, whose value is computed using the (numerically precise) vacancy-state wave function 
amplitudes. By contrast,  the values reported in Ref. \onlinecite{Cazalilla12} are based on scaling arguments 
for the vacancy-induced wave function, which do not take properly into account
the behavior of the vacancy wave function amplitudes along the sites of the system.

Finally, we note that though the results presented for $U$ are for single vacancies, the expression we 
derive is also valid for the multivacancy case. Though an analytical expression for the multivacancy 
wave function case is not available in the literature, those can be obtained numerically and used to 
obtain $U$ in a similar fashion as the one we use here for the single vacancy. 

In summary, we present a full derivation of $U$ for vacancy-induced localized states in graphene systems.
Our results help the understanding of defect-induced carbon magnetism, allowing contact with the 
typical experiments that use samples with billions of atoms, where edges, multivacancies and other 
kinds of disorder are present, which are beyond the reach of other methods, such as DFT

\acknowledgments 
This work has been supported by the Brazilian funding agencies CAPES, CNPq, 
and FAPERJ. 

\appendix
\section{Convergence analysis}
\label{sec:robust}

The tight-binding model calculations are implemented using a supercell of size $M \times N$. 
In the case of graphene nanoribbons, $M$ is determined by the ribbon width, while the longitudinal 
length $N$ has to be conveniently chosen since the system is periodic in this direction. We optimize 
our calculations by choosing the smallest value of $N$ for which $U$ becomes independent (within
less than 1\%) of $N$.

\begin{figure}[h!]
\begin{center}
\includegraphics[width=0.9\linewidth]{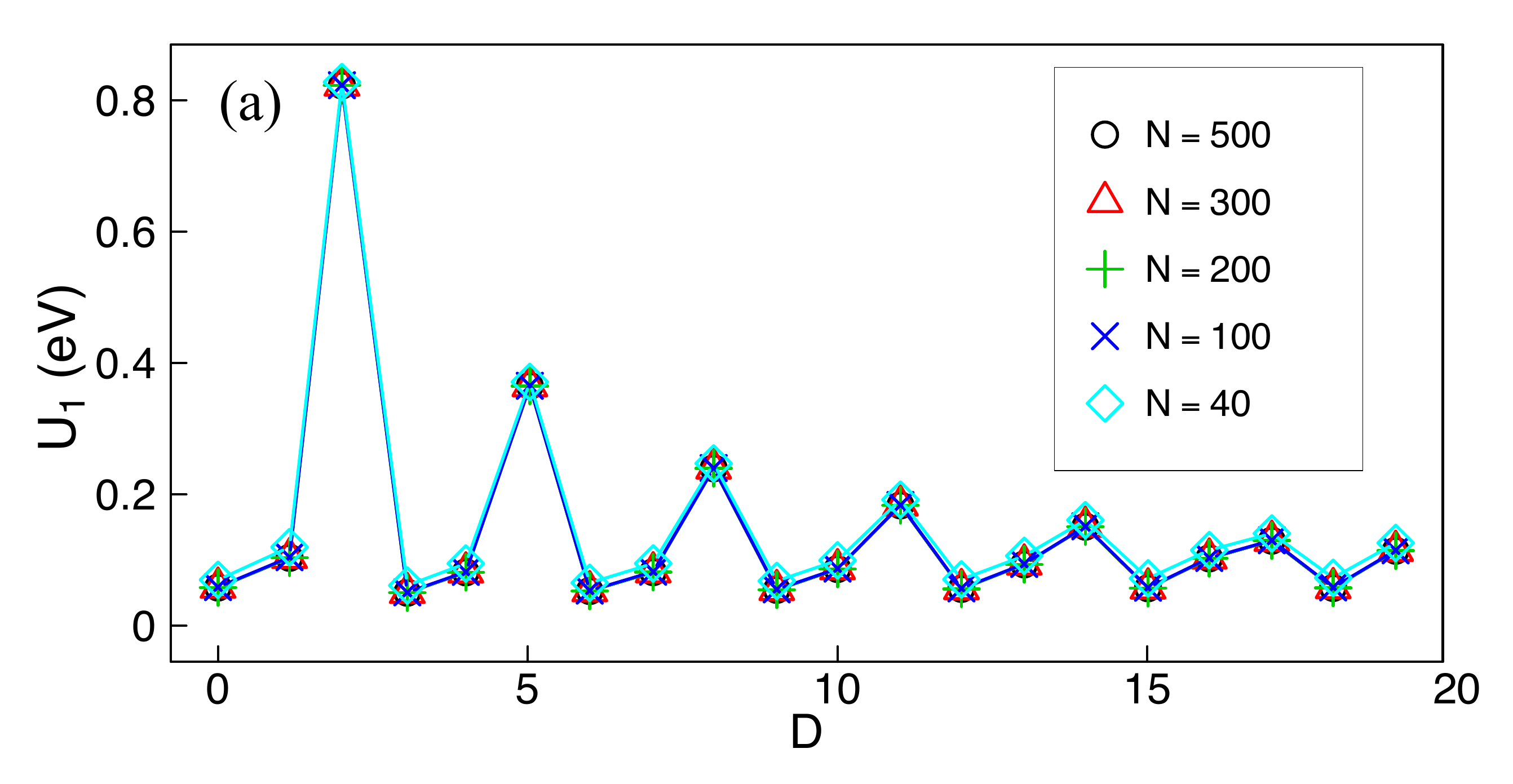}
\includegraphics[width=0.9\linewidth]{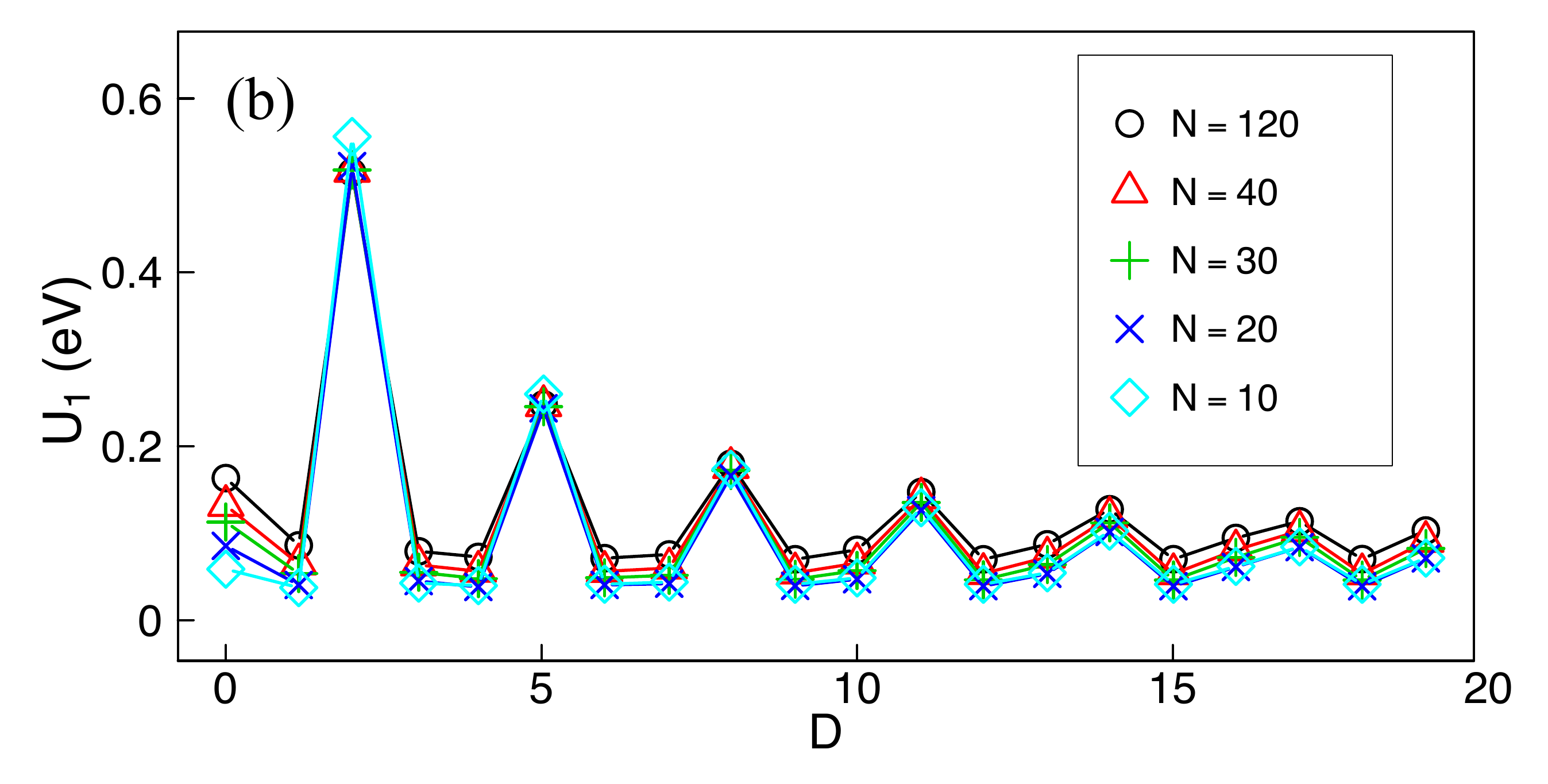}
\vskip-0.2cm
\caption{(Color online) Charging energy as a function of distance for armchair ribbons with length $M=40$ and with varying "infinite length" $N$. (a) Results obtained from the analytical expression of the vacancy wave function. (b) Results from the tight binding model.
\label{fig:Uapendice}}
\end{center}
\end{figure}

In Fig. \ref{fig:Uapendice} we display the behavior of the $U_{1}$ as a function of edge distance $D$ 
for $M=40$ and varying $N$. The results are obtained by using the analytical wave function 
(Fig. \ref{fig:Uapendice}a) and the tight binding model (Fig. \ref{fig:Uapendice}b). 
For $M=40$ we perform tight-binding simulations for $10\le N \le 120$. For larger system sizes
this method becomes computationally intensive and it is advantageous to use analytical wave 
functions. The calculations reveal that our estimates are almost independent of $N$ and show that 
the results already converge for modest values of $N$. These calculations also indicate that the 
discrepancies observed in the estimates of $U_1$ obtained from the two approaches is not an 
artifact of the supercell sizes we use.

%



\end{document}